\documentstyle[aps,amssymb,epsfig,amsmath]{revtex}

\begin{document}
\draft
\title{Spinodal Decomposition in a Binary Polymer Mixture:\\
Dynamic Self Consistent Field Theory and Monte Carlo Simulations}
\author{Ellen Reister, Marcus M\"uller, and Kurt Binder}
\address{Institut f\"ur Physik, WA 331, Johannes Gutenberg Universit\"at, D-55099 Mainz, Germany}
\maketitle
\begin{abstract}
We investigate how the dynamics of a single chain influences the kinetics of early
stage phase separation in a symmetric binary polymer mixture. We consider 
quenches from the disordered phase into the region of spinodal instability. 
On a mean field level we approach this problem with two methods:
a dynamical extension of the self consistent field theory
for Gaussian chains, with the density variables evolving in time, and the method of the
external potential dynamics where the effective external fields are propagated 
in time. Different wave vector dependencies of the kinetic coefficient are 
taken into account.
These early stages of spinodal decomposition are also studied through Monte 
Carlo simulations employing the bond fluctuation model that maps the chains 
-- in our case with 64 effective segments -- on a coarse grained lattice.
The results obtained through self consistent field calculations and  
Monte Carlo simulations can be compared because the time, length, and temperature scales
are mapped onto each other through the diffusion constant, the chain extension,
and the energy of mixing. The quantitative comparison
of the relaxation rate of the global structure factor shows that a kinetic coefficient 
according to the Rouse
model gives a much better agreement than a local, i.e.~wave vector independent, 
kinetic factor. Including fluctuations in the self consistent field calculations 
leads to a shorter time span of spinodal behaviour and a  reduction of the 
relaxation rate for smaller wave vectors and prevents the relaxation rate from 
becoming negative for larger values of the wave vector. This is also in agreement 
with the simulation results.  
\end{abstract}
\section{Introduction}
Over many years the study of phase transitions in fluid mixtures has become 
an important field because of their omnipresent occurrence in nature and more
importantly because of their technological meaning in creating new materials\cite{Paul,Folkes}:
blending of different species can reduce cost, improve processibility, provide synergy between 
components and allow for recycling. 

Sophisticated analytical techniques -- e.g., self consistent field theory\cite{Helfand,Helfand_2,Noolandi,Matsen}
or P-RISM theory\cite{PRISM} -- exist for calculating the phase behaviour and detailed interfacial 
properties of polymer blends. Because of the large extension of polymer chains  equilibrium properties are well described;
however, the analytical description is much less satisfactory for the dynamics of multi--component systems.

The transition from a completely homogeneous mixture
to an equilibrated two phase system is a very long process consisting of a 
sequence of highly inhomogeneous states. Many different methods to analyse 
this process have been applied including experiments, theory, and computational 
simulations.
After quenching a  polymer blend from the one phase region deep into the two 
phase region spinodal decomposition takes place. Different time regimes are 
recognised during spinodal decomposition. During early stages the amplitudes 
of the concentration fluctuations, that are amplified, are still so small that 
they do not interact with each other.  For later times the local composition 
reaches the equilibrium value of the coexisting phases, and nonlinear interactions
between the fluctuation modes become more and more important. At even later stages
hydrodynamics dominates the coarsening\cite{KBSPIN}.

Spinodal decomposition can be approximately described through time dependent 
Ginzburg--Landau theory, also known as
Cahn--Hilliard--Cook theory\cite{Cahn_1,Cahn_2,Cook}. Many efforts have been 
made to numerically
calculate the time evolution of a mixture in the framework of this 
theory\cite{Kotnis,Chakrabarti,Glotzer}.
These calculations are appropriate to give a first insight into how phase
separation takes place, but are far from describing the system quantitatively 
well. Even though binary polymer blends are an ideal testing bed for these
approaches, earlier Monte Carlo simulations\cite{Sariban_1,Sariban_2,Heermann}
 found rather pronounced deviations from their predictions.
To find a quantitatively better description, it is necessary to go 
beyond Ginzburg--Landau theory. Self consistent field 
theory\cite{Helfand,Helfand_2,Noolandi,Matsen} (SCFT) has proven to be one of the most 
successful descriptions of equilibrium properties of polymer mixtures on a 
mean field level. The idea to use SCFT to develop a dynamical mean field
theory is not new\cite{Fraaije_2,Fraaije_1,Yeung,Hasegawa}, but, although SCFT describes equilibrium properties 
well, usually the influence of the single chain dynamics is neglected due to
computational expenses, again leading to a more qualitative than quantitative
description of the phase separation. In most calculations a simple constant Onsager 
coefficient is used which would be appropriate if the movement of the polymers
were comparable to the movement of point particles. Remembering the connectivity
of polymers it is not likely that this local coupling is sufficient
to lead to a quantitatively correct description of the dynamics and other
nonlocal Onsager coefficients have been proposed\cite{Binder_1,deGennes_pap,Pincus}.

In this study we are interested in the early stages of spinodal decomposition
in an incompressible symmetric binary polymer mixture after a quench from the 
one phase region to the two phase region
paying special attention to the differences in the collective dynamics when
the chains obey local and Rouse dynamics.
We therefore
employ two dynamical versions of  self consistent field theory for polymer 
mixtures. The first method, referred to as dynamic self consistent
field theory\cite{Fraaije_2,Fraaije_1} (DSCFT), uses the free energy 
functional
of the SCFT for an incompressible mixture in terms of the 
local composition. This free energy functional leads to a kinetic equation 
for the time dependent local composition which is integrated numerically.
This method has been applied for studying the
ordering kinetics in block copolymer and surfactant systems\cite{Fraaije_2,Vlimmeren_2}.
We use both a local  and a nonlocal Onsager transport coefficient, as is expected 
for polymers obeying Rouse dynamics. 

The underlying idea of the second method is to express the free energy of  the
investigated  system only in terms of the effective external fields, which are 
thermodynamically conjugated to the composition. This description leads
to an equation of motion for the effective external fields, which is integrated
in time.  
Following Maurits {\em et al.}\cite{Maurits} who derived a similar equation of motion 
neglecting random fluctuations we refer to this  method as the external 
potential dynamics (EPD), which has been shown to  
automatically incorporate Rouse dynamics.

The aim of our study is to explore both methods and to investigate the role of the Onsager
coefficient. To decide whether our results resemble any `real' dynamics a 
comparison with results obtained through other methods is preferable. 
Although many experiments have analysed spinodal decomposition
in binary polymer mixtures only few\cite{Schwahn_1,Schwahn_2,Jinnai} have looked at early stages of 
demixing and the influence of the Onsager coefficient.
Another difficulty in comparing experimental results with our calculations 
lies in the fact that actual demixing is influenced by many factors, for 
example, preparation of the probe,
polydispersity, or strong dynamic asymmetries due to different glass transition temperatures of the two species\cite{Tanaka,Tanaka2}.
Therefore a quantitative mapping between theory and experimental systems is difficult.
To gain a quantitative insight  we compare our mean field calculations  with 
results obtained through Monte Carlo simulations. No parameters have to be 
adjusted for this comparison. In the simulations we employ the bond 
fluctuation model\cite{Carmesin_1,Deutsch} which is well established for 
studying properties of polymer melts. It has been shown that the 
equilibrium properties of a polymer mixture as given by the self consistent
field theory are almost quantitatively reproduced through Monte Carlo 
simulations using the bond fluctuation model\cite{MREV,PRE}.

Our paper is organised in the following way. 
In section \ref{sec:SCFT} we introduce the self consistent field theory for 
a binary polymer mixture, explaining how equilibrium properties can be derived
with this method. Then we introduce general aspects of the dynamics in a 
polymer mixture and show how they can be incorporated in the SCFT leading to
dynamic self consistent field theory. The following section presents another
approach to including dynamics in SCFT regarding the effective external 
fields as the time dependent variable. Implications regarding density 
fluctuations in this description are discussed in detail. Section \ref{sec:MC} serves as a brief
introduction to the performed Monte Carlo simulations and explains how a direct
comparison with SCFT calculations is possible.
In section \ref{sec:results} results of our calculations and simulations are presented
first showing the difference between the use of local and nonlocal coupling, 
then comparing the mean field results with the simulations and finally analysing
the influence of random fluctuations. The paper finishes with a summary. 
 
\section{Self Consistent Field Theory}
\label{sec:SCFT}
We consider an incompressible mixture of A and B polymers consisting of $n$ 
polymers in a volume $V=L_x\!\times\!L_y\times\!L_z$ with periodic boundary conditions. There are $n_A$ polymers of kind A
in the system, for B polymers respectively $n_B=n-n_A$. In SCFT\cite{Helfand,Noolandi,Matsen,Shi}
polymers are modelled as Gaussian chains with the 
end--to--end distance $R_e$. In addition we choose the A and B polymers to 
both have the same number of monomers $N_A=N_B=N$ and to be of the same architecture.
The overall particle density in the system is denoted as $\rho=nN/V$.
The microscopic density $\hat{\phi}_A$ of the A monomers can be defined through
the polymer conformations $\{{\mathbf{r}}_{\alpha}(\tau)\}$ :
\begin{equation} 
\hat{\phi}_A=\frac{N}{\rho}\sum^{n_A}_{i_A=1}\int_{0}^{1} d\!\tau
\delta({\mathbf{r}}-{\mathbf{r}}_{i_A}(\tau))
\label{eq:mikro_dens}
\end{equation}
$0\leq\tau\leq 1$ parameterises the contour of a chain.
For B monomers a similar equation holds. (All the following equations regarding
only A monomers, are equivalently valid for B monomers without mentioning this explicitly.) 
Regarding a repulsion between the two kinds of polymers that is expressed 
through the Flory--Huggins parameter $\chi$, the canonical 
partition function has the following form:
\begin{equation}
\begin{split}
Z\sim
\frac{1}{n_A!n_B!}\int&\left(\prod^{n_A}_{i_A=1}\prod^{n_B}_{i_B=1}
{\mathcal{D}}[{\mathbf{r}}_{i_A}]{\mathcal{D}}[{\mathbf{r}}_{i_B}]
{\mathcal{P}}_A[{\mathbf{r}}_{i_A}]{\mathcal{P}}_B[{\mathbf{r}}_{i_B}]\right)\cdot\\
&\cdot\exp\left[-\rho\int_V d^3{\mathbf{r}}\chi\hat{\phi}_A\hat{\phi}_B\right]
\delta(\hat{\phi}_A+\hat{\phi}_B-1)
\end{split}
\label{eq:part_bulk}
\end{equation}
The functional integral ${\mathcal{D}}$ sums over all possible 
conformations of the chains. ${\mathcal{P}}_A[{\mathbf{r}}]$ is the so 
called Wiener measure
\begin{equation}
{\mathcal{P}}_A[{\mathbf{r}}]\sim\exp\left[-\frac{3}{2R_e^2}
\int_{0}^{1}d\!\tau\left(\frac{d {\mathbf{r}}}{d\tau}\right)^2\right]
\label{eq:Wien_mes}
\end{equation}
which represents the statistical weight of a noninteracting Gaussian 
chain. Repulsive interactions at short interparticle distances reduce
fluctuations of the total density. This is incorporated effectively via 
the incompressibility constraint which is expressed through the $\delta$--function in Eq.(\ref{eq:part_bulk})\cite{COMMENT}.

This defines a system with many interacting polymer chains; its partition function
obviously cannot be analytically solved.  Inserting new auxiliary field variables $\Phi_A$, $\Phi_B$, $W_A$, $W_B$
via a Hubbard--Stratonovich transformation, we can reformulate the many polymer problem in terms of a single polymer
in external fields.
\begin{equation}
Z\sim\int{\mathcal{D}}\Phi_A{\mathcal{D}}W_A{\mathcal{D}}\Phi_B{\mathcal{D}}W_B\,\delta(\Phi_A+\Phi_B-1)\exp\left[-F\left[W_A,W_B,\Phi_A,\Phi_B\right]/k_BT\right]
\label{eq:part_new}
\end{equation}
Thus we have found an expression for the canonical free energy
depending on the new variables and the single chain partition function $Q_A$:
\begin{equation}
\begin{split}
\frac{F\left[W_A,W_B,\Phi_A,\Phi_B\right]}{k_BT}&=
-\frac{\bar{\phi_A}\rho V}{N}\ln\frac{Q_A}{n_A}-\frac{\bar{\phi_B}\rho V}{N}\ln\frac{Q_B}{n_B}\\
&-\frac{\rho}{N}\int_V d^3{\mathbf{r}}(W_A\Phi_A+W_B\Phi_B)+
\frac{\rho}{N}\int_V d^3{\mathbf{r}}\chi N\Phi_A\Phi_B,
\end{split}
\label{eq:free_en}
\end{equation}
$\bar{\phi_A}=Nn_A/\rho V$ denoting the average density of the A 
polymers in the system and 
\begin{equation}
Q_A=\int{\mathcal{D}}[{\mathbf{r}}_A]{\mathcal{P}}_A[{\mathbf{r}}]
\exp\left[-\int_0^{1}d\tau W_A({\mathbf{r}}(\tau))\right]
\label{eq:si_ch_part}
\end{equation}
being the partition function of a single A chain in the external field $W_A$.
\subsection{Equilibrium}
The functional integral in Eq.(\ref{eq:part_new}) also cannot be calculated explicitly. 
Therefore we employ a saddle point approximation. This means that only the largest contribution to the 
integrand is considered and the integration does not have to be carried out. The saddle
point approximation of Eq.(\ref{eq:part_new}) is equivalent to the minimisation of 
the free energy (\ref{eq:free_en}) with respect to the auxiliary variables. The values
of the fields and densities at the saddle point are denoted by lower case letters and
are given by the set of equations that has to be solved self consistently:
\begin{eqnarray}
\frac{\delta F}{\delta W_A}\arrowvert_{w_A}=0 &\quad:\quad& 
\phi_A=-\frac{\bar{\phi}_A V}{Q_A}\frac{\delta Q_A}{\delta w_A} \equiv \phi^*_A[w_A] \label{eq:saddle_phi_A}\\
\frac{\delta F}{\delta W_B}\arrowvert_{w_B}=0 &\quad:\quad&
\phi_B=-\frac{\bar{\phi}_B V}{Q_B}\frac{\delta Q_B}{\delta w_B} \equiv \phi^*_B[w_B] \label{eq:saddle_phi_B}\\
\frac{\delta F}{\delta \Phi_A}\arrowvert_{\phi_A}-\frac{\delta F}{\delta \Phi_B}\arrowvert_{\phi_B}=0 &\quad:\quad&
w_A-w_B=\chi N(\phi_B-\phi_A)\label{eq:eq_w_A}\\
&\phi_A+\phi_B=1& \label{eq:eq_w_B}
\end{eqnarray}
The averages of the microscopic densities are given by:
$\langle \hat \phi_A \rangle = \phi^*_A$ and $\langle \hat \phi_B \rangle = \phi^*_B$.
Eqs.(\ref{eq:saddle_phi_A}) and (\ref{eq:saddle_phi_B}) make clear that in equilibrium the average microscopic density $\langle \hat \phi_A \rangle$
actually equals the thermal average monomer density of
a single A chain in an external field $w_A$. In other words after starting with a description where we would
have to take all possible interactions of every polymer chain into account, see Eq.(\ref{eq:part_bulk}), we have now found a mean field 
description of the system in which it is sufficient to look at a single chain in an effective external field.

To calculate the monomer densities it is useful to define the end segment distribution $q_A({\mathbf{r}},t)$, which gives the 
probability to find the end of a chain with length $t$ at position ${\mathbf{r}}$ when exposed to a field 
$w_A$:
\begin{equation}
q_A({\mathbf{r}},t)=\int{\mathcal{D}}[{\mathbf{r}}(t)]{\mathcal{P}}_A[{\mathbf{r}}(t)]
\delta({\mathbf{r}}(t)-{\mathbf{r}})\exp\left[-\int^t_0 d\tau w_A({\mathbf{r}}(\tau))\right],
\label{eq:end_seg_dis}
\end{equation} 
For this end segment distribution the following diffusion equation holds\cite{Helfand}
\begin{equation}
\frac{\partial q_A({\mathbf{r}},t)}{\partial t}=\frac{1}{6}R_e^2\nabla^2q_A({\mathbf{r}},t)
-w_A q_A({\mathbf{r}},t)
\label{eq:diff_eq_end}
\end{equation}
with the boundary condition $q_A({\mathbf{r}},0)=1$. After finding a solution to this equation the monomer density  is immediately given through the following
expression :
\begin{equation}
\phi^*_A({\mathbf{r}})=\frac{\bar{\phi}_AV}{Q_A}\int_0^1 dt\,q_A({\mathbf{r}},t)q_A({\mathbf{r}},1\!-\!t)
\label{eq:dens_calc}
\end{equation}
The single chain partition function can be calculated with:
\begin{equation}
Q_A=
\int_V d^3{\mathbf{r}}q_A({\mathbf{r}},1)
\label{eq:si_ch_part_calc}
\end{equation}
After having presented all necessary equations to calculate equilibrium 
properties of polymer mixtures, we are now interested in using this description
for calculating the dynamics in a polymer mixture. Two ways to achieve this are
introduced in the following sections.
\subsection{Dynamic Self Consistent Field Theory (DSCFT)}

Because of the fact that the concentrations of the polymers are conserved the continuity equation is valid:
\begin{equation}
\frac{\partial\phi_A({\mathbf{r}},t)}{\partial t}+\nabla {\mathbf{J}}({\mathbf{r}},t)=0
\label{eq:cont_eq} 
\end{equation}
${\mathbf{J}}({\mathbf{r}},t)$ denoting the current density of the monomers at position ${\mathbf{r}}$ at time $t$.
One now assumes\cite{Binder_1,deGennes_pap,Pincus} a linear relation between the current density and the gradient of the exchange potential
$\mu=\frac{\delta F}{\delta\phi_A}-\frac{\delta F}{\delta\phi_B}$ (Note there is only one independent
chemical potential in the system because of the incompressibility constraint)
\begin{equation}
{\mathbf{J}}({\mathbf{r}},t)=-\int_Vd{\mathbf{r}}^{\prime3}\Lambda({\mathbf{r}},{\mathbf{r}}^\prime)
\nabla\mu({\mathbf{r}}^\prime,t)
\label{eq:Onsager_intro}
\end{equation}
with the kinetic coefficient $\Lambda({\mathbf{r}},{\mathbf{r}}^\prime)$ describing the connection between the 
force acting on the monomers through the gradient of the chemical potential at position ${\mathbf{r}}^\prime$
and the resulting current density at position ${\mathbf{r}}$. This describes a 
purely relaxational dynamics, effects due to hydrodynamic flow are not captured. 
The Onsager coefficient $\Lambda$ can be modelled in different ways: The simplest approach would be local coupling which results in the Onsager coefficient being proportional to the local density. Bearing in 
mind that polymers have a certain extension it is clear that nonlocal coupling should lead to a better
description, although local coupling is often used in calculations of dynamic models 
based on Ginzburg--Landau type energy functionals for simplicity reasons\cite{Kotnis,Chakrabarti,Glotzer}. In the Rouse model forces
acting on a monomer caused by the other monomers are also taken into account\cite{Rouse,Doi}. This leads to a kinetic factor
that is proportional to the pair--correlation function\cite{Binder_1,deGennes_pap,Pincus,Maurits}. 
These two approaches lead to the following Onsager coefficients:
\begin{align}
\Lambda_{\text{local}}(\mathbf{r})&=DN\phi_A({\mathbf{r}},t)
\phi_B({\mathbf{r}},t)&\text{local coupling}\label{eq:local_dyn}\\
\Lambda_{\text{Rouse}}(\mathbf{r},\mathbf{r}^{\prime})&\approx DN{\bar{\phi}_A}{\bar{\phi}_B}P_0({\mathbf{r}},{\mathbf{r}}^{\prime})&\text{Rouse}\label{eq:Rouse_dyn}
\end{align}
$D$ denoting the single chain diffusion constant and $P_0({\mathbf{r}},{\mathbf{r}}^{\prime})$ the 
pair--correlation function. The Rouse Onsager coefficient written in 
Eq.\eqref{eq:Rouse_dyn} 
is only approximately valid, when the pair--correlation 
function is the same for both polymer species. A more general expression is
found in reference\cite{Maurits}.

Another model for nonlocal coupling is the reptation
model\cite{Doi,deGennes_pap_2} which is appropriate for polymer melts with very long chains, i.e.~very entangled chains. Hereby the idea is
that a polymer chain is constrained by the other polymers and is forced to move along the polymer tube 
axis. This dynamics is expected\cite{Binder_1,deGennes_pap,Pincus} to also lead to an Onsager coefficient that is proportional to the pair--correlation 
function. Therefore the influence of single chain Rouse and reptation
dynamics on the collective dynamics of the system is qualitatively comparable.
Reptation shall not be regarded in the further study. 

Eqs.(\ref{eq:cont_eq}) and \eqref{eq:Onsager_intro} lead to the following diffusion equation the last term 
representing noise that obeys the fluctuation--dissipation theorem.
\begin{equation}
\frac{\partial\phi_A({\mathbf{r}},t)}{\partial t}=\nabla\int_Vd{\mathbf{r}}^{\prime3}
\Lambda({\mathbf{r}},{\mathbf{r}}^\prime)\nabla\mu({\mathbf{r}}^\prime,t)+\eta({\mathbf{r}},t)
\label{eq:diff_eq}
\end{equation}
After the Fourier transformation this diffusion equation has a simple form:
\begin{equation}
\frac{\partial\phi_A({\mathbf{q}},t)}{\partial t}=
 -\Lambda({\mathbf{q}})q^2\mu({\mathbf{q}},t)+\eta({\mathbf{q}},t)
\label{eq:diff_eq_q}
\end{equation}

These diffusion equations implicitly assume that the relaxation time of the chain conformations is smaller than the time scale on which
composition fluctuations evolve. The chain conformations are expected to be `in equilibrium' with respect to the instantaneous spatially varying
composition.

To use the introduced diffusion equation (\ref{eq:diff_eq}) in the frame of SCFT, let 
us return to the general expression (\ref{eq:part_new}) that describes the canonical 
partition function depending upon the variables $\Phi_A$, $\Phi_B$, $W_A$, 
$W_B$ that are independent of each other. If we now employ a saddle point 
approximation in the variables $W_A$ and $W_B$ we find Eqs.(\ref{eq:saddle_phi_A}) and 
(\ref{eq:saddle_phi_B}), that describe a unique relation between the fields $w_A$, $w_B$ and the
densities $\phi_A$, $\phi_B$. This means we can calculate the densities 
explicitely with Eqs.(\ref{eq:diff_eq_end}), (\ref{eq:dens_calc}), and 
(\ref{eq:si_ch_part_calc}),  if the fields
$w_A$ and $w_B$ are known. Therefore with this approximation the free energy $F$ 
becomes a function that depends on the densities (or fields) only: 
$F\left[\phi_A,\phi_B,w_A[\phi_A],w_B[\phi_B]\right]=F\left[\phi_A,\phi_B\right]$.
With this free energy it is now possible to calculate the exchange potential 
$\mu({\mathbf{r}})$:
\begin{equation}
\frac{\mu({\mathbf{r}})}{k_BT}=\frac{\delta F[\phi_A({\mathbf{r}}),\phi_B({\mathbf{r}})]}{\delta \phi_A({\mathbf{r}})}-\frac{\delta F[\phi_A({\mathbf{r}}),\phi_B({\mathbf{r}})]}{\delta \phi_B({\mathbf{r}})}
=\frac{1}{N}\left\{\chi N(\phi_B({\mathbf{r}})-\phi_A({\mathbf{r}}))-(w_A[\phi_A({\mathbf{r}})]-w_B[\phi_B({\mathbf{r}})])\right\}
\label{eq:chem_pot} 
\end{equation}

The simple form of diffusion equation (\ref{eq:diff_eq_q}) and the diffusion equation  for the end segment 
distribution of a polymer chain (\ref{eq:diff_eq_end}) suggests to use a Fourier expansion of all spatially dependent variables for actual calculations 
because the Fourier functions are eigenfunctions of the squared gradient 
$\nabla^2$. The following
set of orthonormal functions is used in all our SCFT calculations:
\begin{equation}
\begin{split}
f_{lmn}({\mathbf{r}})=\text{norm}(l)\,\text{norm}&(m)\,\text{norm}(n)\,\cos\left(\frac{2\pi l}{L_x}x\right)\,
                          \cos\left(\frac{2\pi m}{L_y}y\right)\,\cos\left(\frac{2\pi n}{L_z}z\right)\quad\quad l,m,n=0,1,2,\ldots\\
 \text{norm}&(i)=\begin{cases}\sqrt{2} &i\not=0\\
                               1      &i=0
                \end{cases}
\end{split}
\label{eq:Four_exp}
\end{equation}

We have now found all necessary equations to numerically calculate the time 
evolution of the densities in a binary polymer mixture, leading us the 
following procedure we refer to as the dynamic self consistent field theory 
(DSCFT) method:
First we have a given density profile at time $t\!=\!0$. As mentioned before Eqs.(\ref{eq:diff_eq_end}), (\ref{eq:dens_calc}) and (\ref{eq:si_ch_part_calc}) give us the possibility to explicitly calculate
the single chain densities for known external fields. Unfortunately the inversion is needed because we have 
given densities and have to find the matching external fields. This leads to a set of nonlinear equations,
that is numerically solved through the Newton--Broyden method\cite{Broyden}. 
After the external fields, that `create' the given density profile, have been 
found the 
exchange potential $\mu$ is calculated via Eq.(\ref{eq:chem_pot})
and then inserted into the diffusion equation (\ref{eq:diff_eq_q}). The diffusion equation is subsequently integrated
using a simplified Runge--Kutta method. This leads to a new given density after a discrete time step and the 
whole procedure starts anew.

Apart from regarding local dynamics in our DSCFT calculations it is favourable
to consider nonlocal coupling because this leads to a better description of
the dynamics in a polymer mixture. However in DSCFT the difficulty in using
Rouse dynamics lies in the computational expense of calculating the 
pair--correlation function for each time step. As an approximation during
early stages of demixing the pair--correlation function of a homogeneous 
melt, as it is given through the random phase approximation (RPA)\cite{deGennes}, is 
used leading to the following Onsager coefficient:
\begin{equation}
\Lambda(q)=DN\bar{\phi_A}\bar{\phi_B}\frac{2(x+e^{-x}-1)}{x^2}
\end{equation}
$x$ is defined as $x=R_e^2{\mathbf{q}}^2/6$ with $R_e$ denoting the end--to--end 
distance of a polymer.
\subsection{External Potential Dynamics}
\label{sec:EPD}
In the previously introduced DSCFT method we needed a way to reduce the 
number of independent variables in the partition function. Through the saddle 
point approximation in the fields we obtained a free energy functional in
terms of the densities. This, in turn, yields a Langevin equation for the dynamics
of the densities.
For the external potential dynamics (EPD) method\cite{Maurits} we are looking for a way to 
express the dynamics of the binary polymer mixture through an equation of motion 
for the external fields $W_A$ and $W_B$. 
Our starting point is again the canonical partition function (\ref{eq:part_bulk}).
Via a Hubbard--Stratonovich transformation we introduce the the field variables
$W\!=W_A\!-\!W_B$ and $U\!=\!W_A\!+\!W_B$
and obtain:
\begin{eqnarray}
Z &\sim& \frac{1}{n_A!n_B!}\int\left(\prod^{n_A}_{i_A=1}\prod^{n_B}_{i_B=1}
         {\mathcal{D}}[{\mathbf{r}}_{i_A}]{\mathcal{D}}[{\mathbf{r}}_{i_B}]
         {\mathcal{P}}_A[{\mathbf{r}}_{i_A}]{\mathcal{P}}_B[{\mathbf{r}}_{i_B}]\right)\cdot  \nonumber \\
&&       \cdot \exp\left[-\rho\int_V d^3{\mathbf{r}}\frac{\chi}{4} \left\{ (\hat{\phi}_A+\hat{\phi}_B)^2
                                                                        -(\hat{\phi}_A-\hat{\phi}_B)^2 \right\}\right]
        \delta(\hat{\phi}_A+\hat{\phi}_B-1) \nonumber \\
  &\sim& \frac{1}{n_A!n_B!}\int\left(\prod^{n_A}_{i_A=1}\prod^{n_B}_{i_B=1}
         {\mathcal{D}}[{\mathbf{r}}_{i_A}]{\mathcal{D}}[{\mathbf{r}}_{i_B}]
         {\mathcal{P}}_A[{\mathbf{r}}_{i_A}]{\mathcal{P}}_B[{\mathbf{r}}_{i_B}]\right)\cdot  \nonumber \\
&&       \cdot \int {\mathcal{D}}U {\mathcal{D}}W \exp\left[-\frac{\rho V \chi}{4} \right]
               \exp\left[-\frac{\rho}{N}\int_V d^3{\mathbf{r}} \left\{\frac{W}{2}(\hat\phi_A-\hat\phi_B)+\frac{W^2}{4\chi N} \right\}\right]
               \exp\left[-\frac{\rho}{N}\int_V d^3{\mathbf{r}}\; \frac{U}{2}(\hat{\phi}_A+\hat{\phi}_B-1)\right]
	       \label{eq:zUW}
\end{eqnarray}
This defines a free energy function $G$ in terms of the fields $U$ and $W$:
\begin{equation}
{\cal Z}\sim \int{\mathcal{D}}U{\mathcal{D}}W\exp\left[-G\left[U,W\right]/k_BT\right]
\label{eq:part_UW}
\end{equation}
with
\begin{equation}
\frac{G\left[U,W\right]}{k_BT}=
-\frac{\bar{\phi_A}\rho V}{N}\ln\frac{Q_A[(U\!+\!W)/2]}{n_A}-\frac{\bar{\phi_B}\rho V}{N}\ln\frac{Q_B[(U\!-\!W)/2]}{n_B}
+\frac{\rho}{N}\int_V d^3{\mathbf{r}}\left[
\frac{W^2}{4\chi N} -\frac{1}{2}(U-\frac{\chi N}{2})\right]
\label{eq:free_en_UW}
\end{equation}
Alternatively, we could have started with the free energy functional (\ref{eq:part_new}) and integrate out the Gaussian
variables $\Phi_A$ and $\Phi_B$. So far no approximations have been used, but to find an energy functional that 
only depends upon $W$ we now employ a saddle point approximation with respect 
to $U$:
\begin{equation}
\frac{\delta G[U,W]}{\delta U}\arrowvert_{U^*}=0\quad:\quad\phi_A^*({\mathbf{r}})+\phi_B^*({\mathbf{r}})=1
\label{eq:saddle_U}
\end{equation}
Here we use the definition, see Eqs.\eqref{eq:saddle_phi_A} and
\eqref{eq:saddle_phi_B}: $\phi_A^*=-\frac{\bar{\phi}_A V}{Q_A}\frac{\delta Q_A}{\delta W_A}\label{phi1}$. For $\phi_B^*$ the equivalent definition 
applies. In equilibrium the field variable $U\!=\!W_A\!+\!W_B$ is conjugated
to the overall density of the system, which is constant in an incompressible 
mixture. We therefore believe that the influence of this approximation on the
description of the system through the field $W$ is very small. We shall discuss this in detail below.

 If we replace $U$ with the $U^*[W]$ that fulfils this constraint we end up 
with a free energy functional that only depends on the field variable $W$:
\begin{equation}
\frac{G[W({\mathbf{r}})]}{k_B T}=\frac{\rho V \chi}{4}+\frac{\rho}{N}\int_V d^3{\mathbf{r}}\frac{W^2}{4\chi N} 
-\frac{\bar{\phi_A}\rho V}{N}\ln\frac{Q_A[(U^*\!+\!W)/2]}{n_A}-\frac{\bar{\phi_B}\rho V}{N}\ln\frac{Q_B[(U^*\!-\!W)/2]}{n_B}
\label{eq:free_en_W}
\end{equation}
where we have used the fact that  adding a constant field $\xi$ to $U^*$ does not change the value of $G[W({\mathbf{r}})]$.
We chose  $\xi$ in such a way that $\int_Vd^3{\mathbf{r}}\; U^*=0$.

A difficulty in describing the system with the order parameter $W$ is the interpretation of the field fluctuations in terms of the physical
density fluctuations. We can calculate the averages of the microscopic densities {\em after} the saddle point integration over $U$. These
averages are marked by the subscript EPD. To this end, we introduce a local exchange potential $\Delta \mu$, which couples to the microscopic 
density difference $\hat \phi_A - \hat \phi_B$:
\begin{eqnarray}
\tilde{\cal Z}[\Delta \mu]  &\sim& \frac{1}{n_A!n_B!}\int\left(\prod^{n_A}_{i_A=1}\prod^{n_B}_{i_B=1}
         {\mathcal{D}}[{\mathbf{r}}_{i_A}]{\mathcal{D}}[{\mathbf{r}}_{i_B}]
         {\mathcal{P}}_A[{\mathbf{r}}_{i_A}]{\mathcal{P}}_B[{\mathbf{r}}_{i_B}]\right)
	 \exp\left[-\frac{\rho}{N}\int_V d^3{\mathbf{r}}  \left\{\frac{\Delta \mu}{2}(\hat\phi_A-\hat\phi_B) \right\}\right]
	 \cdot  \nonumber \\
&&       \cdot \int {\mathcal{D}}W \exp\left[-\frac{\rho V \chi}{4} \right]
         \exp\left[-\frac{\rho}{N}\int_V d^3{\mathbf{r}} \left\{\frac{W}{2}(\hat\phi_A-\hat\phi_B)+\frac{W^2}{4\chi N} \right\}\right]
         \exp\left[-\frac{\rho}{N}\int_V d^3{\mathbf{r}}\; \frac{U^*[W]}{2}(\hat{\phi}_A+\hat{\phi}_B-1)\right] \nonumber \\
&\sim&  \int {\mathcal{D}}W \exp\left[ -\tilde{G}[W,\Delta \mu] \right] \qquad \mbox{with} \nonumber \\
\tilde{G}[W,\Delta \mu] &=& \frac{\rho V \chi}{4} + \frac{\rho}{N} \int_V {d}^3{\bf r}\; \frac{W^2}{4\chi N}
-\frac{\bar\phi_A\rho V}{N}\ln\frac{Q_A[(U^*+W+\Delta\mu)/2]}{n_A}-\frac{\bar\phi_B\rho V}{N}\ln\frac{Q_B[(U^*-W-\Delta\mu)/2]}{n_B}\label{eq:av_1}
\end{eqnarray}
Thermodynamic averages of the microscopic density difference are obtained via functional derivatives:
\begin{eqnarray}
\langle \hat\phi_A({\bf r})-\hat\phi_B({\bf r}) \rangle_{\rm EPD}
                        &=&  -\frac{2N}{\rho}\frac{1}{\tilde{\cal Z}[\Delta \mu]} \frac{\delta\tilde{\cal Z}[\Delta \mu]}
			                                        {\delta \Delta \mu({\bf r})}\arrowvert_{\Delta \mu = 0}  \nonumber \\
                        &=& \langle \phi^*_A({\bf r})- \phi^*_B({\bf r}) \rangle
\label{eq:phi_av_ok} \\
\langle [\hat\phi_A({\bf r})-\hat\phi_B({\bf r})][\hat\phi_A({\bf r'})-\hat\phi_B({\bf r'})] \rangle_{\rm EPD}
                        &=&  \left(\frac{2N}{\rho}\right)^2 \frac{1}{\tilde{\cal Z}[\Delta \mu]} \frac{\delta^2\tilde{\cal Z}[\Delta \mu]}
                                          {\delta \Delta \mu({\bf r})  \delta\Delta \mu({\bf r'})}\arrowvert_{\Delta \mu = 0}  \nonumber \\
                        &=& \left\langle
                         [\phi^*_A({\bf r})-\phi^*_B({\bf r})][\phi^*_A({\bf r'})-\phi^*_B({\bf r'})]  \right\rangle
	 -\frac{N}{\rho}\left\langle\frac{\delta \phi_A^*({\bf r})}{\delta W_A({\bf r'})}+\frac{\delta \phi_B^*({\bf r})}{\delta W_B({\bf r'})}
			\right\rangle
\label{eq:phi^2_av_ok}
\end{eqnarray}

Similarly, we can calculate the fluctuations of the total density, which are induced by the saddle point approximation.
\begin{eqnarray}
\tilde{\cal Z}[ \mu]  &\sim& \frac{1}{n_A!n_B!}\int\left(\prod^{n_A}_{i_A=1}\prod^{n_B}_{i_B=1}
         {\mathcal{D}}[{\mathbf{r}}_{i_A}]{\mathcal{D}}[{\mathbf{r}}_{i_B}]
         {\mathcal{P}}_A[{\mathbf{r}}_{i_A}]{\mathcal{P}}_B[{\mathbf{r}}_{i_B}]\right)
	 \exp\left[-\frac{\rho}{N}\int_V d^3{\mathbf{r}}  \left\{\frac{\mu}{2}(\hat\phi_A+\hat\phi_B) \right\}\right]
	 \cdot  \nonumber \\
&&       \cdot \int {\mathcal{D}}W \exp\left[-\frac{\rho V \chi}{4} \right]
         \exp\left[-\frac{\rho}{N}\int_V d^3{\mathbf{r}} \left\{\frac{W}{2}(\hat\phi_A-\hat\phi_B)+\frac{W^2}{4\chi N} \right\}\right]
         \exp\left[-\frac{\rho}{N}\int_V d^3{\mathbf{r}}\; \frac{U^*[W]}{2}(\hat{\phi}_A+\hat{\phi}_B-1)\right] \nonumber \\
&\sim&  \int {\mathcal{D}}W \exp\left[ -\tilde{G}[W,\mu \right] \qquad \mbox{with} \nonumber \\
\tilde{G}[W,\mu] &=& \frac{\rho V \chi}{4} + \frac{\rho}{N} \int_V {d}^3{\bf r}\; \frac{W^2}{4\chi N}
-\frac{\bar\phi_A\rho V}{N_A}\ln\frac{Q_A[(U^*+W+\mu)/2]}{n_A}-\frac{\bar\phi_B\rho V}{N_B}\ln\frac{Q_B[(U^*-W+\mu)/2]}{n_B}
\end{eqnarray}
Moments of the total density averaged over the field configurations of $W$ are given by:
\begin{eqnarray}
-\frac{\rho}{2N} \langle \hat\phi_A({\bf r})+\hat\phi_B({\bf r}) \rangle_{\rm EPD}
                        &=& \frac{1}{\tilde{\cal Z}[\mu]} \frac{\delta\tilde{\cal Z}[\mu]}
			                                        {\delta \mu({\bf r})}\arrowvert_{ \mu = 0}  \nonumber \\
\Rightarrow \qquad \qquad \langle \hat\phi_A({\bf r})+\hat\phi_B({\bf r}) \rangle_{\rm EPD} 
                        &=&  \langle \phi^*_A({\bf r}) + \phi^*_B({\bf r}) \rangle = 1\\
\left(\frac{\rho}{2N}\right)^2 \langle (\hat\phi_A({\bf r})+\hat\phi_B({\bf r})(\hat\phi_A({\bf r'})+\hat\phi_B({\bf r'}) \rangle_{\rm EPD}
                        &=& \frac{1}{\tilde{\cal Z}[\mu]} \frac{\delta^2\tilde{\cal Z}[ \mu]}
                                          {\delta\mu({\bf r})  \delta \mu({\bf r'})}\arrowvert_{ \mu = 0}  \nonumber \\
\Rightarrow \qquad \qquad  \langle (\hat\phi_A({\bf r})+\hat\phi_B({\bf r})(\hat\phi_A({\bf r'})+\hat\phi_B({\bf r'}) \rangle_{\rm EPD}
&=& 1 - \frac{N}{\rho}\left\langle
			 \frac{\delta \phi_A^*({\bf r})}{\delta W_A({\bf r'})}+\frac{\delta \phi_B^*({\bf r})}{\delta W_B({\bf r'})}
			\right\rangle \label{eq:dens}
\end{eqnarray}
These equations describe the actual fluctuations of the microscopic composition of the system {\em after} the saddle point approximation,
i.e., of the EPD method. Having performed a saddle point integration in $U$, we have ignored fluctuations, and we cannot expect Eqs.(\ref{eq:phi^2_av_ok})
or (\ref{eq:dens}) to be accurate. Indeed, while the incompressibility constraint is fulfilled {\em on average} Eq.(\ref{eq:dens}) demonstrates that
the saddle point approximation leads to spurious fluctuations of the total density. In Appendix \ref{app:RPA_EPD} we use the random phase approximation (RPA)
to evaluate Eq.(\ref{eq:phi^2_av_ok}) and show explicitely the deviations between Eq.(\ref{eq:phi^2_av_ok}) and the well known RPA--structure factor.

Alternatively, we can deduce the exact averages  from the full free energy 
functional $G[U,W]$ in Eq.(\ref{eq:free_en_UW}) by introducing a local exchange potential $\Delta \mu$ like in Eq.\eqref{eq:av_1}.
After a variable substitution $W\!+\!\Delta\mu\!\rightarrow\!W$ this leads to:
\begin{equation}
\tilde{G}[U,W,\Delta \mu ] = G[U,W] - \frac{\rho}{N}  \int_V d^3{\mathbf{r}}\; \frac{-\Delta \mu^2 + 2 \Delta \mu W}{4\chi N}
\end{equation}
With this free energy functional we obtain the exact averages of the microscopic densities:
\begin{eqnarray}
\langle \hat\phi_A({\bf r})-\hat\phi_B({\bf r}) \rangle_{UW}
&=& - \frac{1}{\chi N}\langle W \rangle_{UW} 
\label{eq:phi_av} \\
\langle [\hat\phi_A({\bf r})-\hat\phi_B({\bf r})][\hat\phi_A({\bf r'})-\hat\phi_B({\bf r'})] \rangle_{UW}
        &=&  \frac{-2 \delta({\bf r}-{\bf r'})}{\rho \chi } +   \frac{1}{(\chi N)^2} \langle W({\bf r})W({\bf r'})\rangle_{UW}               
	\label{eq:phi^2_av}
\end{eqnarray}
Recently, Ganesan and Fredrickson\cite{PREPRINT} have used a complex Langevin method to sample the fluctuations of both fields $U$ and $W$ and
have obtained the average $A$--monomer density as $\langle \phi^*_A \rangle_{UW}$. Since $U$ has to be complex to make the last term in 
Eq.(\ref{eq:zUW}) a proper representation of $\delta(\hat \phi_A + \hat \phi_B-1)$ individual contributions to this average also have
an imaginary part and the numerical procedure is quite involved.

We expect the saddle point integration over $U$ to be accurate whenever $G[U,W]$ can be well approximated by a parabola in $U-U^*$. 
In this case the fluctuations of $W$ are only very little affected by the saddle point approximation in $U$ and the fluctuations of $W$ in the EPD method will
closely mimic the fluctuations of $W$ of the exact partition function (\ref{eq:zUW}). Hence, we can use  Eq.(\ref{eq:phi^2_av}) with
$\langle W({\bf r})W({\bf r'})\rangle_{UW} \approx  \langle W({\bf r})W({\bf r'})\rangle$ (i.e.,  {\em after} the saddle point approximation)
to obtain a very good approximation for the structure factor. In Appendix \ref{app:RPA_EPD} we confirm that in RPA the fluctuations of $W$ are, of course,
not affected by the saddle point integration over $U$. Therefore we use  the Fourier transform of Eqs.(\ref{eq:phi_av}) and (\ref{eq:phi^2_av}) in our calculations:
\begin{eqnarray}
\langle\phi_A(q)-\phi_B(q)\rangle&=&-\frac{1}{\chi N}\langle W(q)\rangle\label{eq:phi_av_Four}\\
\langle|\phi_A(q)-\phi_B(q)|^2\rangle&=&-\frac{2}{\rho V\chi}+\frac{1}{(\chi N)^2}\langle |W(q)|^2\rangle\label{eq:phi^2_av_Four}
\end{eqnarray}

Because the composition is conserved and $\langle \hat\phi_A-\hat \phi_B\rangle \sim \langle W \rangle$,
we expect the order parameter $W$ with which we are now describing our system to also be a conserved quantity. Therefore the dynamics 
of $W$ are given through the relaxational dynamics of a model B system, referring to the classification introduced by
Hohenberg and Halperin\cite{Hohenberg}.
\begin{equation}
\frac{\partial W({\mathbf{r}})}{\partial t}=\nabla_{{\mathbf{r}}}
\int_V\Lambda({\mathbf{r}},{\mathbf{r}}^{\prime})\nabla_{{\mathbf{r}}^\prime}
\mu_W({\mathbf{r}}^\prime)+\eta({\mathbf{r}},t)
\label{eq:diff_eq_EPD}
\end{equation}
with the chemical potential being the first derivative of the free energy with
respect to the order parameter:
\begin{equation}
\mu_W({\mathbf{r}})=\frac{\delta G[W({\mathbf{r}})]}{\delta W({\mathbf{r}})}
=\frac{1}{N2\chi N}\left(W+\chi N\left[\phi_A^*({\mathbf{r}})-\phi_B^*({\mathbf{r}})\right]\right)
\label{eq:chem_pot_EPD}
\end{equation}
The Fourier transform of this new diffusion equation is:
\begin{equation}
\frac{\partial W(q)}{\partial t}=-\Lambda(q)q^2\frac{1}{2N\chi N}
\left[W(q)+\chi N(\phi_A^*(q)-\phi_B^*(q))\right]+\eta(q)
\label{eq:diff_eq_EPD_q}
\end{equation}
$\eta$ is white noise that obeys the fluctuation--dissipation theorem.
The method using this diffusion equation we refer to as the EPD method\cite{Maurits}.

We have found a diffusion equation that describes the dynamics in 
terms of the external field $W=W_A-W_B$ and leads to the right physical
equilibrium. A similar equation without noise has been derived by Maurits {\em et al.}\cite{Maurits}.
The question to be asked is whether this dynamics represents 
any actual physical dynamics and how the choice of the Onsager coefficient
influences the dynamics of the densities. It can be shown\cite{Maurits}, see 
also appendix
\ref{app:Rouse_EPD}, that using local coupling in the EPD method is a good approximation
for Rouse dynamics. The Onsager coefficient that we would have to use in
DSCFT to reproduce Rouse dynamics is given in Eq.(\ref{eq:Rouse_dyn}). The equivalent
(local) kinetic coefficient in the EPD method for the same dynamics of the
densities is:
\begin{equation}
\Lambda_{\text{EPD}}=-2\chi N D
\label{eq:Onsager_EPD}
\end{equation}

For the EPD calculations again the Fourier expansion of Eq.(\ref{eq:Four_exp}) is
used. After having found the initial fields that create the given densities with the
methods used for the DSCFT the chemical potential $\mu_W$ is calculated according to 
Eq.(\ref{eq:chem_pot_EPD}). $\mu_W$ is then plugged into Eq.(\ref{eq:diff_eq_EPD_q}) to 
find the time derivative of the difference in the fields. 
Thereafter $\frac{\partial W}{\partial t}=\frac{\partial W_{A}}{\partial t}-\frac{\partial W_{B}}{\partial t}$ is 
integrated via the simplified Runge--Kutta method. After we have found the new $W=W_A-W_B$, we need to find the variable 
$U^*$ to make sure the incompressibility constraint $\phi_A^*+\phi_B^*=1$ given
through the saddle point approximation (\ref{eq:saddle_U}) 
is fulfilled again using the Newton--Broyden method. The new fields lead us to the 
new chemical potential to calculate 
$\frac{\partial W}{\partial t}=\frac{\partial W_{A}}{\partial t}-\frac{\partial W_{B}}{\partial t}$ and so forth. 

The method of the EPD has two main advantages compared to DSCFT: First of all it 
incorporates nonlocal coupling and secondly proves to be up to an order of magnitude
computationally faster. There are two main reasons for this huge speed up:
In EPD the number of equations that have to be solved via the Newton--Broyden method
to fulfil incompressibility
is just the number of Fourier functions used. The number of equations in DSCFT that have
to be solved
to find the new fields after integrating the densities is twice as large. On the 
other hand, comparing diffusion equation (\ref{eq:diff_eq}) used in the DSCFT method with equation 
(\ref{eq:diff_eq_EPD}) in EPD it is easily seen, that the right hand side of the latter is 
a simple multiplication with the squared wave vector of the relevant mode, whereas
the right hand side of Eq.(\ref{eq:diff_eq}) is a complicated multiplication of three
spatially dependent variables.

\section{Monte Carlo Simulations}
\label{sec:MC}
\subsection{Bond Fluctuation Model}
The Monte Carlo simulations presented in this study make use of the bond fluctuation
model (BFM)\cite{Carmesin_1,Deutsch}, which is a coarse grained lattice model, that incorporates the relevant
features of polymers. These are connectivity of the monomers along a chain, 
excluded volume of the segments and thermal interaction between monomers. In this 
model each effective monomer occupies a cube of the lattice and blocks the eight sites at the cube corners for other monomers. Monomers of a chain are connected by one of 
108 possible bond vectors of length 2, $\sqrt{5}$, $\sqrt{6}$, 3, or $\sqrt{10}$ 
measured in units of the lattice spacing. (All following lengths are assumed to be
given in these units unless an explicit unit is given.) These bond vectors are 
chosen to ensure the excluded volume condition, that makes sure that they do not cross 
each other during their movement. This large number of possible bond 
vectors allows 87 different bond angles, that provide a good approximation for 
continuous connectivity between the monomers of the chain.
Each of the effective monomers represents 3--5 real chemical repeat units\cite{Paul_1,Paul_2}. The 
number density of the occupied sites is chosen to be $\rho=1/16$, which reproduces the 
properties of a polymer melt well. Interactions between the monomers
are modelled through a square well potential with monomers of the same kind 
attracting and monomers of different kinds repelling each other. The interactions 
are chosen to be symmetric and to act inside a radius that extends over the first peak of the
pair correlation function. This means the interactions act up to a distance of
$\sqrt{6}$ which is equivalent to the 54 neighbouring cubes of a monomer.
\begin{equation}
-\epsilon_{AA}= -\epsilon_{BB}= \epsilon_{AB}= 
\begin{cases}
k_BT\epsilon\geqslant0& \text{for}\ r\leqslant\sqrt{6}\\
0&                     \text{for}\ r>\sqrt{6}
\end{cases}
\label{eq:MC_pot}
\end{equation}

The moves used to simulate a purely diffusive movement of the monomers are local 
random monomer hopping moves, where one tries to move a randomly picked monomer to 
a neighbouring lattice site. 
\subsection{Comparison between SCFT Calculations and Monte Carlo Simulations}
The Monte Carlo simulations were carried out with chains of 64 effective segments 
in a box with length $L=160$. With an overall number density of $\rho=1/16$ there are
256 000 particles in the system. 
In the Monte Carlo simulations however the 
choice of $N=64$, which is equivalent to a polymerisation of 200--300 in real 
polymers, is a compromise between the possibility to compare the simulation results
with mean field results and the largest still sensibly manageable amount of 
computational resources, because an increase in the polymerisation leads to both an 
increase in the length scale and a slowing down in the kinetics of phase 
separation.
To actually compare Monte Carlo simulations with SCFT calculations the parameters
of both models have to mapped onto each other:

In SCFT the only present 
length scale is the end--to--end distance $R_e$ of the polymers. It can be measured 
directly in the Monte Carlo simulations $R_e=\sqrt{N}b=25.12$. This gives us the 
length of the system to be $L=160=6.35 R_e$.

In the SCFT calculations only the combination $\chi N$ of Flory Huggins parameter and
chain length enters. This sets the temperature scale. The Flory Huggins parameter $\chi$
can be calculated from the interaction 
parameter $\epsilon$ of the square well potential defining the interaction between monomers in 
the Monte Carlo simulations with the following relation\cite{Muller_1}:
\begin{equation}
\chi=\frac{1}{k_BT}z_{\text{eff}}[\epsilon_{AB}-\frac{1}{2}(\epsilon_{AA}+\epsilon_{BB})]
=2z_{\text{eff}}\epsilon
\end{equation}
$z_{\text{eff}}$ is the effective coordination number in the bulk, i.e.~the average number 
of intermolecular contacts per monomer. We hereby speak of contacts, if the distance
between the monomers is smaller than $\sqrt{6}$. 

The average composition $\bar \phi_A$ is a parameter of the SCFT, the total number density $\rho$
of monomers is only required if fluctuations are considered.

The single chain diffusion constant $D$, which can be extracted from the Monte Carlo 
simulations by measuring the mean square displacements of the chains, gives the 
time scale $\tau=R_e^2/D=1.5\times 10^7$ Monte Carlo steps (MCS). $\tau$ is constant, because $D$ and $R_e$
are almost independent of time and composition\cite{MDYN}.

The length of the chains used in the Monte Carlo simulations is $N=64$. This is
somewhat larger than the entanglement length  $N_e\approx 32$\cite{Paul_1,Paul_2}, therefore, we
are in a cross over regime between Rouse dynamics and reptation\cite{THORSTEN,JOJO}. This means, when we
are comparing dynamic mean field results with Monte Carlo results we expect to find a 
reasonable agreement if we regard Rouse dynamics.

This identifies all parameters of the SCFT calculations (without noise). If we neglect fluctuations, systems 
with identical $\chi N$, $R_e$, and composition but different degree of interdigitation $\rho R_e^3/N$\cite{MREV,Albano} and 
statistical segment length $b=R_e^2/N$ give identical results. The degree of interdigitation controls the 
strength of fluctuations and mean field theory is believed to  be correct in the limit $\rho R_e^3/N \to \infty$.
The statistical segment length sets the smallest length scale the Gaussian description of polymers is valid. If
we were interested in the structure on smaller length scales we would have to use a different chain model (e.g., worm--like 
chain\cite{WORM,WORMF}).

\section{Results}
\label{sec:results}
In the following sections we regard the early stages of spinodal 
decomposition after a quench from the one phase region with $\chi N=0.314$ into the
miscibility gap with $\chi N=5$ for a symmetric binary polymer mixture. 
For much larger incompatibility the width of the interface becomes of the
order of the statistical segment length, and properties on this length scale cannot
be described by the Gaussian chain model. For smaller incompatibilities --in the vicinity of the
critical point $\chi N = 2$-- composition fluctuations are very strong (i.e., non--Gaussian) and the mean field approximation becomes worse.
We will 
first show some general results for the dynamical mean field theory. We will then 
carry on to compare these results with Monte Carlo simulations showing what role the
Onsager coefficient plays. After having neglected random statistical fluctuations
in SCFT so far we will explain how fluctuations are implemented into the two dynamic
mean field theories and discuss their influence on the dynamics.
\subsection{General aspects of spinodal decomposition}
\label{sec:Gen_asp}
\begin{figure}
\begin{center}
\epsfig{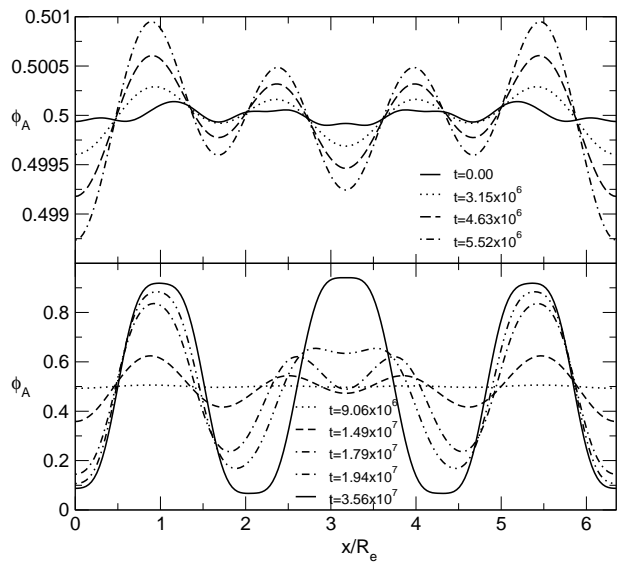}
\end{center}
\caption{\label{fig:fig_1}Density profiles at different times during demixing
after a quench from $\chi N=0.314$ to $\chi N=5$ in a one dimensional system using 12 eigenfunctions. The upper panel shows 
the early stages when concentration fluctuations with wavelengths between a 
third and a quarter of the system size are amplified. In the lower panel later
stages are displayed -- the concentration inside a domain slowly saturates 
leading to sharp interfaces between the coexisting phases. Note the change of scale on the
composition axis between the two panels.}
\end{figure}

If a system is quenched from the one into the two phase region 
the linearized theory of spinodal decomposition\cite{Cahn_1,Cahn_2,Cook} predicts that 
fluctuations with
wave lengths $\lambda$ larger than a critical value $\lambda_c$, i.e wave vectors 
$q$ below a critical value $q_c$, start growing spontaneously. This is illustrated 
in Fig.\ref{fig:fig_1} showing results obtained through DSCFT with local coupling in a one
dimensional system. 
Before the quench at time $t=0$ we have a homogeneous mixture 
with random statistical density fluctuations. After the quench these fluctuations
are amplified. It is apparent that the resulting density profile is well describable
with plane waves, so that the set of basis functions we have chosen proves to be not
only technically convenient for the calculations but also describes the physical
phenomena well. At later times we see that the mode with the wavelength about a 
third of the system size is amplified most until at even later stages a saturation
inside domains takes place until these domains are separated by sharp interfaces.
Later stages of demixing are not appropriately described within our purely diffusive model because
hydrodynamic mechanisms and random fluctuations are neglected but play a
dominant role.
 
\begin{figure}
\begin{center}
\epsfig{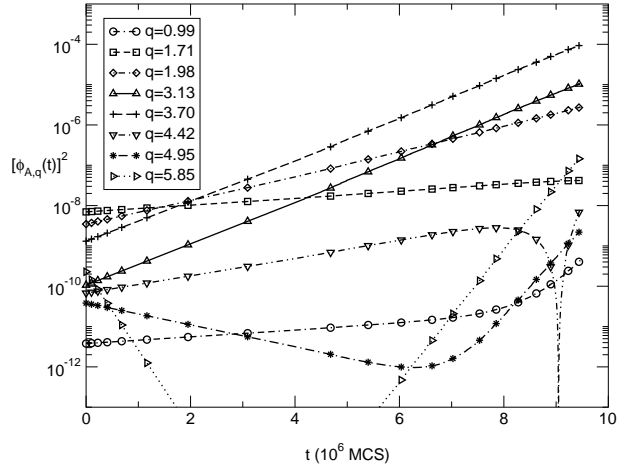}
\end{center}
\caption{\label{fig:phi_An_log_t} Several density modes displayed on a 
logarithmic scale versus time. The displayed values of $q$ are given in
units of $1/R_e$. The results were obtained through DSCFT 
calculations in a three dimensional system of length $L_x\!=\!L_y\!=\!L_z\!=\!6.35 R_e$ 
using $7\!\times\!7\!\times\!7$ functions
for a quench from $\chi N=0.314$ to $\chi N=5$. The expected exponential behaviour during early
stages of demixing is well reproduced.}
\end{figure}

\begin{figure}
\begin{center}
\epsfig{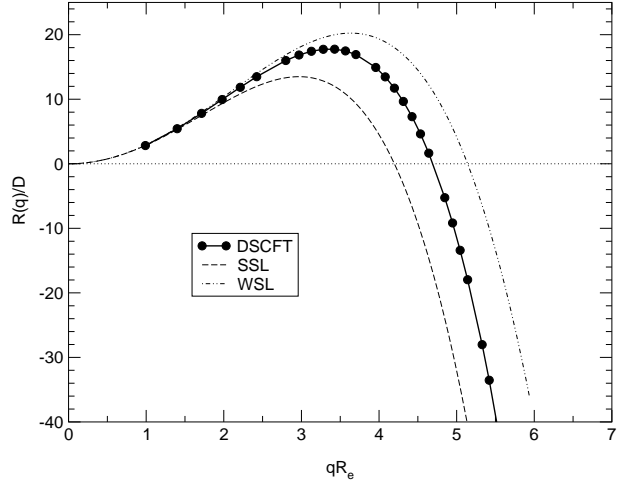}
\end{center}
\caption{\label{fig:Relax_W_S}Corresponding relaxation rate to the density 
modes displayed in Fig.\ref{fig:phi_An_log_t}. Below a critical wave vector 
$q_c$ the density modes are increased spontaneously. 
Modes with larger wave vectors are damped. As expected the results are found to be between the two limits of weak (WSL) and strong segregation (SSL) as given by the Cahn--Hilliard--Cook theory.}
\end{figure}

\begin{figure}
\begin{center}
\epsfig{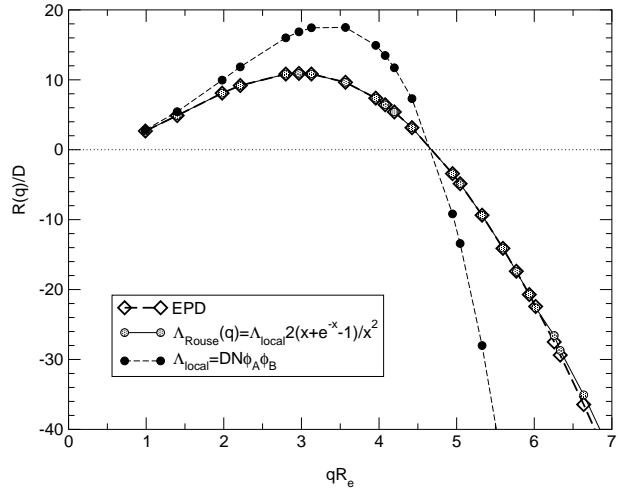}
\end{center}
\caption{\label{fig:EPD}Relaxation rates obtained through  two dimensional DSCFT calculations
using local  and Rouse dynamics and EPD calculations. The DSCFT results using
the pair--correlation function of a homogeneous melt and the EPD results are in
good agreement. }
\end{figure}

Linearised Cahn--Hilliard--Cook theory\cite{Cahn_1,Cahn_2,Cook} predicts an exponential behaviour for the density 
modes $\phi_A({\mathbf{q}})\sim e^{R(q)t}$ as long as the difference $\phi_A({\mathbf{r}})-\bar{\phi}_A$ 
between the actual density and the average density is small. Starting with a 
homogeneous mixture before the quench this requirement is fulfilled during early
demixing. Our density coefficients $\phi_A({\mathbf{q}})$ are equivalent to the 
actual, but discretised, density modes. In Fig.\ref{fig:phi_An_log_t} some of the coefficients
$\phi_A({\mathbf{q}})$ are displayed versus time.
These results were obtained for a three dimensional system using DSCFT with a local 
Onsager coefficient. The exponential behaviour is obviously well reproduced for 
early times. For larger values of $q$ the exponential behaviour changes 
earlier. This is also in agreement with experimental results\cite{Jinnai_2}. 
The faster growth of density modes with smaller wave vectors leads to the 
creation
of small domains on a length scale $\sim 1/q$. These domains then cause density
modes with wave lengths smaller than the extension of the domain to be damped.
 Fig.\ref{fig:Relax_W_S} shows the corresponding relaxation rate $R(q)$ versus
$q$. 
We see that for wave 
vectors with a value below $q_c$ an exponential growth of the density modes sets 
in whereas for $q>q_c$ initial fluctuations are damped. In the growth region there 
is a maximum. The two dashed lines framing our results are the relaxation rates for
temperatures just below the critical temperature (weak segregation limit WSL) and
for very low temperatures (strong segregation limit SSL) as they are given by the 
Cahn--Hilliard--Cook theory:
\begin{equation}
\frac{R(q)}{q^2}=-2\Lambda(0)[\chi_S(\bar{\phi_A}))-\chi]\left(1-\frac{q^2}{q_c^2}\right)
\end{equation}
with
\begin{equation}
\begin{split}
q_c&=\frac{\sqrt{2k}}{R_e}\sqrt{\bar{\phi_A}(1-\bar{\phi_A})}[\chi N-\chi_S(\bar{\phi_A})N]^{1/2}\\
&=\frac{\sqrt{k}}{R_e}(\chi N/2-1)^{1/2}
\end{split}
\end{equation}
The last line is only valid for symmetric mixtures with 
$\bar{\phi_A}=\bar{\phi_B}=0.5$ when the spinodal point lies at 
$\chi_SN=2$. $k$ is found 
to be 18 in the WSL\cite{deGennes_pap} and 12 in the SSL\cite{Roe}. The maximum growth rate is found for $q_m=q_c/\sqrt{2}$.  With $\chi N=5$ lying 
between the SSL and the WSL our result was to be expected.
The influence of Rouse dynamics on the relaxation rate is displayed in
Fig.\ref{fig:EPD}. Here we have presented results for a two dimensional system using DSCFT with Onsager coefficients describing both local kinetics and Rouse 
kinetics for an approximately homogeneous mixture and EPD.
The EPD and the DSCFT results using Rouse dynamics are virtually identical, but 
remembering that the EPD method is computationally much more favourable, this method
should be used when Rouse kinetics is considered. Local and nonlocal coupling
lead to distinct differences in the relaxation rates: in the range $0\leq q\leq q_c$ nonlocal coupling leads to a reduction of the relaxation rate and a 
shift of the maximum to smaller values of the wave vector. Also in the region 
$q>q_c$ the initial density fluctuations are not as strongly damped as for 
local dynamics. 

Note that the difference in growth rate between the local dynamics in the WSL and SSL is very similar
in magnitude to the difference in growth rate between the full SCFT calculation using a local or a Rouse Onsager 
coefficient. This demonstrates that the square gradient expression for the free energy is not sufficiently accurate:
quantitative deviations from experiments or simulations might be either due to the additional approximations of the 
square gradient approach or the wave vector dependence of the Onsager coefficient. To compare simulations and theory quantitatively
and to extract the interplay between single chain dynamics and the kinetics of collective composition fluctuations a
quantitative self consistent field calculation is required.

\subsection{Comparison between Monte Carlo results and dynamic mean field results}
The Monte Carlo simulations were performed on a Cray T3E using a trivial 
parallelisation scheme running 64 configurations in parallel to achieve good
statistics. 5 400 000 MCS were performed which is equivalent to
45 days of CPU time per processor.
The EPD and DSCFT calculations, the results of which are presented in this 
section, are for a three dimensional system using $7\!\times\!7\!\times\!7=343$ basis 
functions.   The equivalent to
12 700 000 MCS were performed taking approximately 65 days on a Cray J90.

\begin{figure}[htbp]
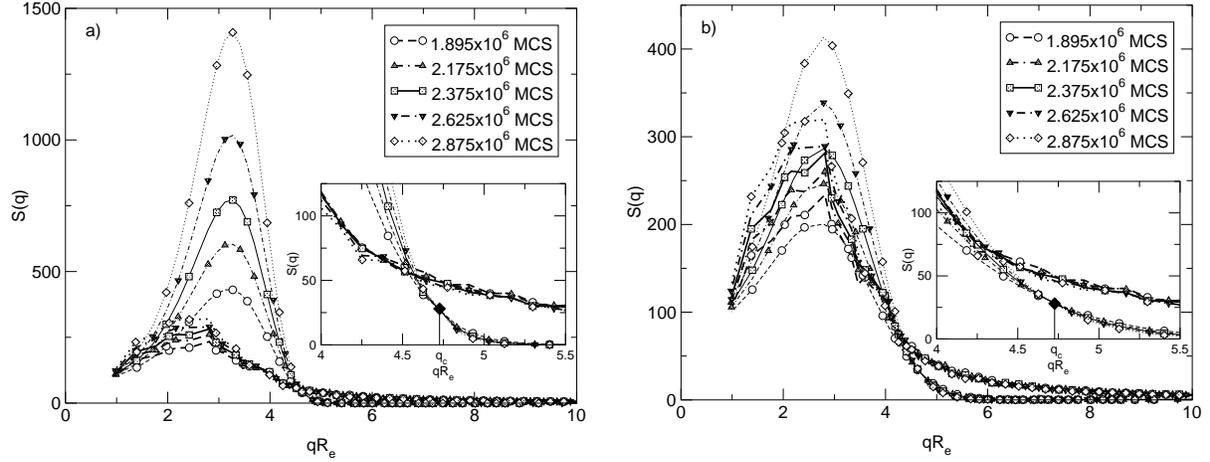

\begin{center}
\begin{minipage}{0.45\textwidth}
\epsfig{file=fig_5a.eps,width=0.95\textwidth,clip=}\\
\end{minipage}
\begin{minipage}{0.45\textwidth}
\epsfig{file=fig_5b.eps,width=0.95\textwidth,clip=}\\
\end{minipage}
\end{center}
\caption{\label{fig:Glostr_SCF_MC}Global structure factor versus wave vector
for different times. Broader lines represent Monte Carlo results,
thin lines with the same symbols the corresponding DSCFT results.  Panel a) compares DSCFT with local coupling with the Monte Carlo simulations. Local dynamics obviously overestimates the growth rate and shifts the maximum growth rate to larger values. Panel b) compares Rouses dynamics with Monte Carlo results showing better agreement. The inset graphs show the behaviour of the global structure
factor in the area of $q_c$: While the mean field results lead to a common 
intersection point defining $q_c$, the Monte Carlo lines do not cross each
other in a single point making the definition of $q_c$ impossible.}
\end{figure}

\begin{figure}
\begin{center}
\epsfig{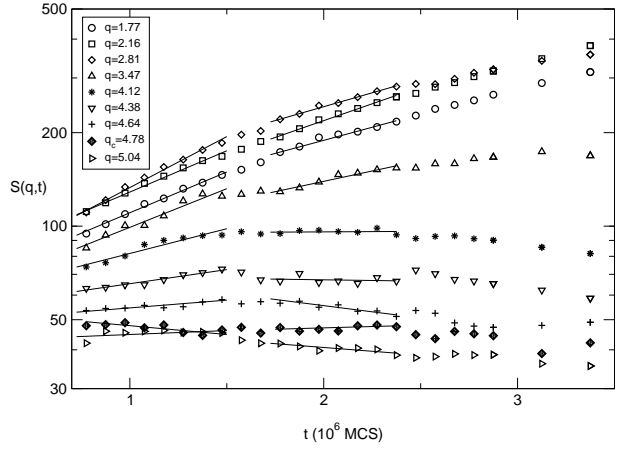}
\end{center}
\caption{\label{fig:MC_Sq_t_log} Global structure factor versus time for a few
randomly chosen values of $q$. The values of the wave vectors are given in 
units of $1/R_e$. $q_c$ corresponds to the critical wave vector extracted 
from DSCFT. An exponential growth of the modes is found. Modes with a 
smaller growth rate change their exponential behaviour earlier than those
growing faster.}  
\end{figure}

\begin{figure}
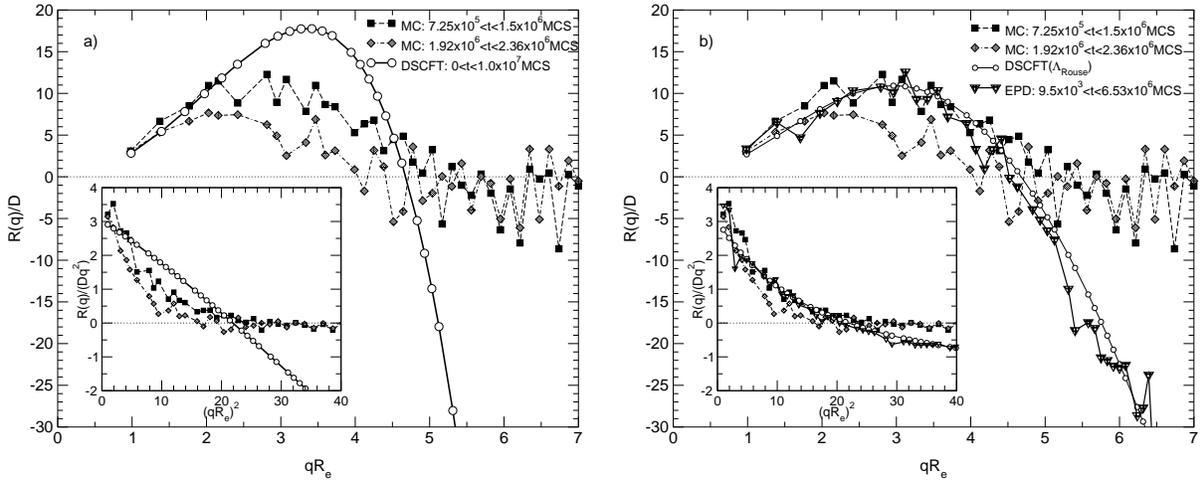

\begin{center}
\begin{minipage}{0.45\textwidth}
\epsfig{file=fig_7a.eps,width=0.95\textwidth,clip=}\\
\end{minipage}
\begin{minipage}{0.45\textwidth}
\epsfig{file=fig_7b.eps,width=0.95\textwidth,clip=}\\
\end{minipage}
\end{center}
\caption{\label{fig:Relaxationsrate_local_neu}Corresponding relaxation rates to
Fig.\ref{fig:Glostr_SCF_MC}. Panel a) compares the Monte Carlo relaxation 
rates with DSCFT calculations with local dynamics. Panel b) compares Monte Carlo results with EPD and DSCFT calculations with Rouse dynamics. 
For earlier 
times good agreement in the growth region is found but Monte Carlo simulations
show an earlier change in the exponential behaviour. The inset figures are the
corresponding Cahn plots, where one displays $R(q)/q^2$ versus $q^2$.} 
\end{figure}

The global structure factor defined by
\begin{equation}
S(q,t)=\left\langle\int_Vd^3{\mathbf{r}} \left|[\phi_A({\mathbf{r}},t)-\phi_B({\mathbf{r}},t)]
e^{i\mathbf{qr}}\right|^2\right\rangle
\end{equation} 
is an important experimentally measurable quantity for the description
of the phase separation process.
 In Fig.\ref{fig:Glostr_SCF_MC} the global structure factor is plotted
versus  wave vector $q$ for different times, thin lines
denoting SCFT results,
wide lines with the same symbols corresponding Monte Carlo simulation results. (Because the SCFT calculations could only
be performed for a single starting configuration instead of 64 like in the
Monte Carlo simulations, for the initial time the global structure factor
of a homogeneous mixture given by RPA\cite{deGennes} was used. Global 
structure factors for later times were extracted through the exponential time 
dependence of the density modes in the SCFT calculations.) 
 Part a) compares Monte Carlo with local kinetics 
and part b) Monte Carlo with Rouse kinetics. Fig.\ref{fig:Glostr_SCF_MC}a) clearly shows no quantitative
agreement between DSCFT with local kinetics and the Monte Carlo simulations: The peak in 
SCFT grows much quicker and also the position of the peak is too far
right. As has been mentioned before, including Rouse dynamics leads to a reduction of 
the relaxation rate in the growth region and the wave vector with the maximum growth 
rate is shifted towards a smaller value. Fig.\ref{fig:Glostr_SCF_MC}b) satisfies these expectations:
The position of the peaks for SCFT and Monte Carlo  almost coincide and the growth rates of the 
peak are much closer to each other although SCFT still overestimates the growth rate. A
more detailed comparison is possible if we plot the corresponding relaxation rates
versus wave vector. As we have seen before the modes of the density calculated
with the dynamic mean field theory follow a clear exponential behaviour. To 
derive the relaxation rates of the global structure factor, we need to plot
the modes of the global structure factor on a logarithmic scale versus time. 
This is done in Fig.\ref{fig:MC_Sq_t_log} for some randomly chosen values of
$q$. The modes with a q value smaller than the critical value $q_c$ derived 
through DSCFT show a clear exponential behaviour. This 
behaviour changes with time, especially the bigger the value of $q$ of the
growing mode the sooner the change in the exponential behaviour takes place.
This is qualitatively in agreement with the results we obtained through dynamic
mean field theory, see section \ref{sec:Gen_asp}. Modes with $q\gtrsim q_c$ are more or less
constant in our figure. 
Therefore an estimation of $q_c$ from the Monte Carlo results alone would suffer from
some ambiguity: of course, this is not only because the accuracy of the Monte Carlo results
is limited, but is a matter of principle\cite{KBSPIN}. The interplay of fluctuations and nonlinear
effects has the consequence that a well--defined $q_c$ does not exist.
For times earlier than the displayed interval random 
fluctuations influence the behaviour of the modes so strongly, that an
exponential behaviour is not visible. 
The relaxation rates  resulting from the fits to the points in Fig.\ref{fig:MC_Sq_t_log} for the two different time intervals indicated through the solid lines and the corresponding mean field relaxation rates are displayed in Fig.\ref{fig:Relaxationsrate_local_neu}. Only the exponential behaviour of the earlier
of the two marked time intervals may be interpreted as the expected behaviour 
of early spinodal decomposition. For later times a gradual change away from
the exponential behaviour sets in. The second fit  shows an apparent
exponential behaviour, because the time interval is too small to resolve the 
change. Hence the resulting `relaxation rate' for the later time interval 
may only be treated as an
indication for the deviation from the earlier exponential behaviour.      
Part a)
shows the relaxation rate for DSCFT with local dynamics in the time interval 
$0\leq t\leq 1\times 10^7 \text{MCS}$ and the Monte Carlo results for the two time intervals
$7.25\times 10^5\leq t\leq 1.5\times 10^6 \text{MCS}$ and $1.92\times 10^6\leq t\leq 2.63\times 10^6 \text{MCS}$; part b) shows the 
same results but regarding Rouse dynamics through the use of EPD and DSCFT with
Rouse dynamics. The inset figures are the corresponding Cahn plots displaying
$R(q)/q^2$ versus $q^2$.
As is obvious from the
behaviour of the global structure factor, local dynamics gives a relaxation rate that is
much too large and the wave vector of the maximal growth rate is also too big.
The Cahn plot of the mean field relaxation rate shows the linear
behaviour, which is expected for local coupling. This is in strong disagreement with the Monte Carlo results. 
For Rouse dynamics we actually find an almost quantitative agreement for earlier times in the 
region of the positive growth rate.  For later times, however, the relaxation rate 
decreases in the Monte Carlo simulations -- this has also been observed in 
experimental studies\cite{Schwahn_1} --, while SCFT still shows the same exponential
behaviour as is found for earlier times. In the Cahn plot we see that Rouse 
dynamics leads to a nonlinear run of $R(q)/q^2$ versus $q^2$. This is 
obviously also the case for the simulation results and is in agreement
with earlier simulations\cite{Sariban_1,Sariban_2,Heermann} and experimental observations\cite{Schwahn_1,Schwahn_2,Jinnai}.
This nonlinear behaviour of the Cahn plot is related to the fact that we consider a deep quench,
for which $q_c R_e >1$\cite{Binder_1}. For a shallow quench, for which $\chi N$ exceeds the critical value $\chi_c N=2$\cite{deGennes}
only slightly, one has $q_c R_e<1$\cite{Binder_1}, and then the theory would yield a linear Cahn plot, also consistent with corresponding
observations\cite{KBSPIN}. The latter case is less interesting, however, because then the polymer mixture is to a large extent 
equivalent in behaviour to a fluid mixture of small molecules, and there is no longer an effect of internal Rouse relaxation modes on the phase separation 
dynamics in this limit.

If we now compare our results for larger wave vectors 
 we see big discrepancies independent of the chosen dynamics in SCFT: At 
$q_c$ the structure factor is independent of time in the early stages of the SCFT calculations, 
i.e.~structure factors at different times exhibit a common crossing point at $q_c$, see the inset graphs of Fig.\ref{fig:Glostr_SCF_MC}. No
such intersection occurs in the simulation data. Mean field theory damps the density modes 
with $q>q_c$ while Monte Carlo simulations lead to a relaxation rate
fluctuating around zero. This behaviour is also seen in the global structure factor, see
Fig.\ref{fig:Glostr_SCF_MC}:  The right side of the SCFT peak decays fast, 
while the Monte Carlo peak
is much broader with a slower decay.
Both the earlier change of the exponential behaviour and the form of the relaxation rate
for larger wave vectors are the result of random fluctuations. As was expected, the 
influence of fluctuations during the very early stages of spinodal decomposition on the
growth of the density modes is rather small, but they are crucial for smaller wave 
lengths and determine the change in the exponential behaviour, because random 
fluctuations cause some modes to reach an amplitude where the non--linear regime
sets in early. 

\subsection{The influence of random fluctuations} 
The diffusion equations we have used so far are completely deterministic, but 
obviously in all dynamic processes random statistical fluctuations are present.
To regard these fluctuations we have to add a random force $\eta$ to our 
diffusion equations (\ref{eq:diff_eq}) and (\ref{eq:diff_eq_EPD}) that is linked to the Onsager 
coefficient through the fluctuation--dissipation theorem:
\begin{equation}
\begin{split}
\langle\eta({\mathbf{r}})\rangle&=0\\
\langle\eta({\mathbf{r}},t)\eta({\mathbf{r}}^{\prime},t^{\prime})\rangle&=2k_B T
\Lambda({\mathbf{r}},{\mathbf{r}}^{\prime})\nabla^2
\delta({\mathbf{r}}-{\mathbf{r}}^{\prime})\,\delta(t-t^{\prime})
\end{split}
\label{eq:fluct_diss}
\end{equation}
or Fourier transformed:
\begin{equation}
\begin{split}
\langle\eta_q\rangle&=0\\
\langle\eta_q(t)\eta_{-q}(t^{\prime})\rangle&=\langle|\eta_q|^2\rangle=2k_B T
\Lambda(q)q^2\delta(t-t^{\prime})
\end{split}
\end{equation}
In our calculations the diffusion equation is integrated through discrete time
intervals $\delta t$ that are determined through the Runge--Kutta scheme. In
DSCFT, for example, we use the Langevin equation:
\begin{equation}
\phi_{A,q}(t+\delta t)=\phi_{A,q}(t)+\delta t \Lambda(q)q^2\mu_q+f_q(\delta t)
\end{equation}
with $f_q(\delta t)$ expressing random fluctuations that obey the fluctuation
dissipation theorem\cite{Vlimmeren_1,Gardiner}:
\begin{equation}
f_q(\delta t)=\sqrt{2\Lambda(q)q^2}\sqrt{\delta t}\ r
\label{eq:rand_fluct}
\end{equation}
$r$ is a random number with the properties $\langle r\rangle=0$, 
$\langle r^2\rangle=1$. 

The analogue $f_q(\delta t)$ is used to perform EPD calculations with 
fluctuations. The difficulty in this case is to interpret the resulting 
fluctuations of the external field $W\!=\!W_A\!-\!W_B$ in terms of the density 
$\Phi\!=\!(\phi_A\!-\!\phi_B)/2$. As we have seen in section \ref{sec:EPD} the field and the
density fluctuations are linked to each other through equations 
\eqref{eq:phi_av_Four} and \eqref{eq:phi^2_av_Four}.

\begin{figure}
\begin{center}
\epsfig{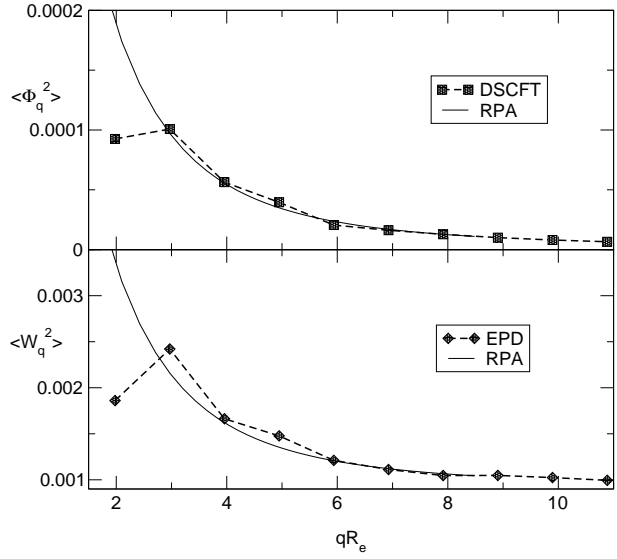}
\end{center}
\caption{\label{fig:phi-w_RPA}The average $\langle|\Phi_q|^2\rangle$ derived
with DSCFT and 
$\langle |W_q|^2\rangle$ derived with EPD displayed versus $q$. Both 
calculations are valid for a homogeneous, $\chi N=1.8$, one dimensional system
with length $L=6.35 R_e$. For larger values of $q$ good agreement with the
RPA averages is found. For smaller values of $q$ too few independent 
configurations of the system were taken into account.}
\end{figure}

\begin{figure}
\begin{center}
\epsfig{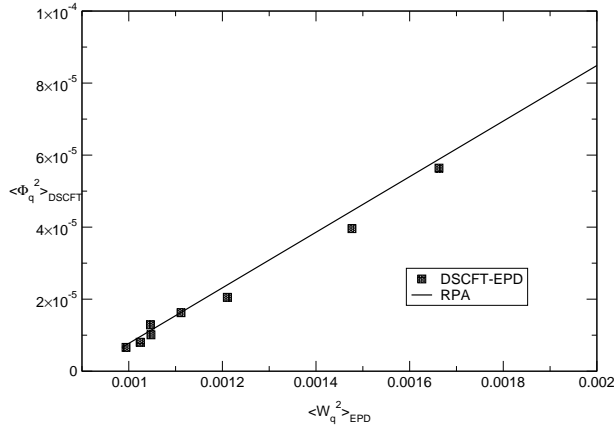}
\end{center}
\caption{\label{fig:EPD_DSCFT_FLUCT}The linear behaviour expressed in 
Eq.\eqref{eq:phi^2_av_Four} is well reproduced, as we can see, when 
$\langle|\Phi_q|^2\rangle$, obtained through DSCFT, is displayed versus 
$\langle |W_q|^2\rangle$, obtained through EPD. The points to the right 
correspond to small wave vectors, those on the left to large wave vectors. The
solid line is the corresponding RPA result.}
\end{figure}

To ensure that the way the fluctuations are included is correct and the 
validity of these Eqs.\eqref{eq:phi_av_Four} and \eqref{eq:phi^2_av_Four}
is given, we consider a 
homogeneous, $\chi N=1.8$, one dimensional system with length $L_x=6.35 R_e$ using
both DSCFT and EPD. From the DSCFT calculations we derive 
$\langle|\Phi_q|^2\rangle$ for each wave vector while the EPD method leads to
$\langle |W_q|^2\rangle$. In both cases we  averaged over 10000  snapshots  of the density or the field made during their time evolution. 
In Fig.\ref{fig:phi-w_RPA}  $\langle|\Phi_q|^2\rangle$ and  $\langle |W_q|^2\rangle$ are plotted versus $q$. The solid line is the expected result from
RPA. For larger values of $q$ we find good agreement, but the amount of used
configurations was not enough to find reliable results for smaller values of $q$ because the correlation times are much longer due to the diffusive dynamics.
If we plot $\langle|\Phi_q|^2\rangle$ versus 
$\langle |W_q|^2\rangle$, as is done in Fig.\ref{fig:EPD_DSCFT_FLUCT}, we 
find good agreement of our data with the expected linear behaviour. 

\begin{figure}
\begin{center}
\epsfig{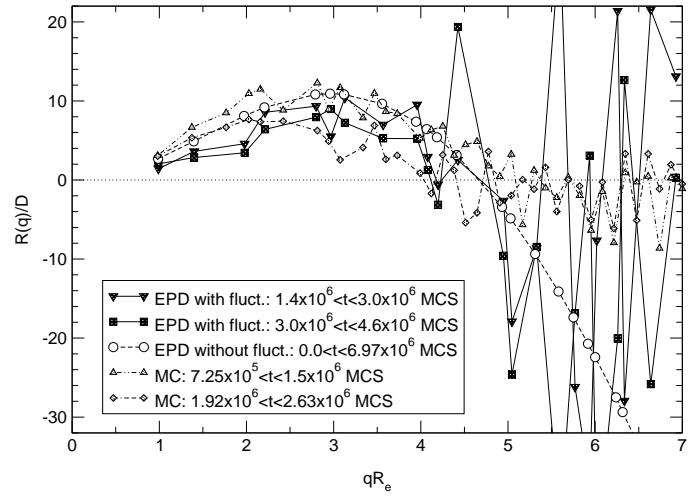}
\end{center}
\caption{\label{fig:Relaxationsrate_dw_dt_fluct}Relaxation rates obtained 
through Monte Carlo simulations and EPD calculations in two dimensions
with random fluctuations
for different time intervals. Fluctuations lead in both methods to an earlier
change in the exponential behaviour of the increasing modes.}
\end{figure}

After having proven that the above treatment of random fluctuations 
leads to the expected behaviour we focus on the influence of fluctuations
on spinodal decomposition. Monte Carlo results show that the exponential
behaviour changes earlier than in SCFT calculations without fluctuations 
resulting in a reduced relaxation rate. The relaxation rates for different time
intervals obtained through the Monte Carlo simulations and the EPD method are
presented in Fig.\ref{fig:Relaxationsrate_dw_dt_fluct}.
The EPD results were obtained by averaging the time evolution of the fields of
64 two dimensional configurations. For earlier time intervals the relaxation rate is 
quantitatively very similar to the rate without fluctuations in the region 
below $q_c$. For later times however when calculations without fluctuations 
still show the same exponential behaviour the relaxation rate is reduced as is 
also seen in the Monte Carlo simulations. In the range above $q_c$ the modes of
the fields or densities do not follow an exponential behaviour but fluctuate
around zero. Consequently the relaxation rate is not well defined in this
region leading to strong fluctuations of the relaxation rate in
Fig.\ref{fig:Relaxationsrate_dw_dt_fluct}. For very early times when the density 
modes are still of the order of the fluctuations of the homogeneous system it 
is also not possible to see an exponential behaviour because the density changes
caused through random fluctuations conceal the growth of the modes during very  early spinodal 
decomposition.
\section{Summary}
In this study we analysed the influence of single chain dynamics on the 
collective diffusion  during early stages of spinodal decomposition in a 
symmetric binary polymer blend. We used the SCFT for polymer mixtures to
explore two versions of dynamical mean field theory. The single chain dynamics
enters  these descriptions through an Onsager coefficient. The first  method
we call DSCFT propagates the densities in time and gives us the possibility
to model both local dynamics and  approximately the nonlocal dynamics  we 
expect for the Rouse model. In DSCFT the correct treatment of Rouse
dynamics would involve the calculation of the pair--correlation function at every
time step, which is a computationally rather expensive task. On the other hand
during early stages of demixing the mixture is only weakly inhomogeneous
so that the use of the pair--correlation function of a homogeneous mixture, which
is analytically known, serves as a sufficient approximation. In the second 
method the instantaneous configuration is not described by the densities but by the effective 
external field (EPD). We find  a Langevin equation for the external  field which
using `local' kinetics in the fields is found to describe a polymer mixture
with Rouse dynamics. Apart from `automatically' including Rouse dynamics this
EPD method has the big advantage of being up to an order of magnitude computationally faster
than the DSCFT method.

First numerical calculations with these methods for a quench from
the one phase, $\chi N=0.314$, to the two phase, $\chi N=5$, region neglecting 
random fluctuations show a clear exponential behaviour of the density modes,
as was expected for a mean field description. The relaxation rate
of the density mode is strongly influenced by the choice of the Onsager 
coefficient: In the growth region Rouse dynamics reduces
the relaxation rate compared to local dynamics and the position of the maximum
growth rate is shifted to smaller values of $q$. For $q>q_c$ however Rouse 
dynamics causes the modes to be damped less quickly. We also find good 
agreement for early stages between the DSCFT method using the pair--correlation 
function for the homogeneous system and the EPD method.

To quantitatively test these mean field predictions we compare them
with results obtained through Monte Carlo simulations  employing the bond 
fluctuation model. The chains  used in these simulations are expected to show 
Rouse behaviour because of the chosen chain length $N=64$. The comparison is possible
without any adjustable parameter.
We compare the global structure factor, which is the experimentally accessible
quantity, and the relaxation rates. Local dynamics in DSCFT overestimates the
growth of the global structure factor by far, but the agreement is better
for Rouse dynamics especially for earlier times. Neglecting random 
fluctuations in our mean field calculations proves to be justified for wave
vectors with positive relaxation rates and earlier times, but should be 
included to investigate later times. 

Fluctuations can easily be included in DSCFT and EPD. The difficulty in the
EPD method is the fact that the field fluctuations have to be interpreted
in terms of the physical density fluctuations. We can find a relation between
the field and the density fluctuations. EPD calculations with fluctuations
lead to an earlier change in the exponential behaviour of the density modes 
as was also the case for the Monte Carlo simulations. The missing dampening
of the modes with $q>q_c$ as is found in the simulations is also reproduced.

We have seen that the single chain dynamics has a pronounced influence on the 
collective dynamics of a polymer mixture. Comparing quantitatively Monte Carlo 
simulations and dynamical mean field theory we have validated the mean field 
calculations. Note, however, that we have considered a deep quench far below the critical point;
for shallow quenches close to the critical point mean field theories are not expected to be accurate.
The later stages of spinodal decomposition are not accessible
with either method. During later times hydrodynamical interactions become important.
Lattice model Monte Carlo simulations lack a hydrodynamical mechanism. In dynamical 
mean field calculations hydrodynamic coupling can be included\cite{Maurits_2,Koga}, and,
hence, they can be extended beyond the validity of the lattice model.

\subsection*{Acknowledgement}
We have benefitted from discussion with D. D{\"u}chs, F. Schmid, and V. Ganesan.
Financial support was provided by the Graduierten Kolleg ``supramolecular systems'' of the university of Mainz and the DFG under grant Bi314/17 
in the priority program ``wetting and structure formation at interfaces''. Generous access to computers at the NIC J{\"u}lich and the HLR Stuttgart
are gratefully acknowledged.

\appendix
\section{Representation of Rouse dynamics through EPD}
\label{app:Rouse_EPD}
As mentioned before 
it is possible to show  that the EPD method using local coupling is a good 
approximation for reproducing Rouse dynamics of the physical densities.
The derivation of the EPD method we present in this section was introduced by
N.M. Maurits {\em et al.}~and can be found in 
reference\cite{Maurits}.

Following the method we used to include dynamics in the SCFT to achieve the DSCFT
method we employ a saddle point approximation in the external fields which 
leads to a bijective relation between the external fields $w_A$, $w_B$ and
the densities $\phi_A$, $\phi_B$. This means we can choose with which of the two variable 
sets we would like to calculate.
On the other hand the pair--correlation function which is part of the Rouse  
Onsager coefficient can be expressed as the functional derivative of the
density with respect to the external potential, see also Eq.(\ref{eq:pair-corr}).
\begin{equation}
 \frac{\delta\phi_A({\mathbf{r}})}{\delta w_A(\mathbf{r'})}=-\frac{\bar{\phi}_A}{N}P_{0}(\mathbf{r,r'})
\label{eq:pair-corr_EPD}
\end{equation}
To calculate in $w$ space we have to transform the time derivative of the densities according to the chain 
rule:
\begin{equation}
\frac{\partial\phi_A({\mathbf{r}},t)}{\partial t}= \int\frac{\delta\phi_A({\mathbf{r}},t)}{\delta w_A({\mathbf{r}}^{\prime},t)}\frac{\partial w_A({\mathbf{r}}^{\prime},t)}{\partial t}d{\mathbf{r}}^{\prime 3}=-\frac{\bar{\phi}_A}{N}\int P_{0}({\mathbf{r,r'}})\frac{\partial w_A({\mathbf{r}}^{\prime},t)}{\partial t}d{\mathbf{r}}^{\prime 3}
\end{equation}
Combining this equation with the diffusion equation for Rouse dynamics (\ref{eq:Rouse_dyn}) this leads us to
\begin{equation}
-\frac{\bar{\phi}_A}{N}\int P_{0}({\mathbf{r,r'}})\frac{\partial w_A({\mathbf{r}}^{\prime},t)}{\partial t}d{\mathbf{r}}^{\prime 3}
= D\frac{\bar{\phi}_A}{N}\nabla_{\mathbf{r}}\int_VP_{0}({\mathbf{r}},{\mathbf{r}}^{\prime})\nabla_{{\mathbf{r}}^{\prime}}\mu_A({\mathbf{r}}^{\prime})d{\mathbf{r}}^{\prime 3}
\end{equation}
Using the approximation 
\begin{equation}
\nabla_{{\mathbf{r}}}P_{0}({\mathbf{r,r'}})\cong-\nabla_{\mathbf{r'}}P_{0}(\mathbf{r,r'})
\label{eq:approx_EPD}
\end{equation}
one easily arrives at 
\begin{equation}
-\int_V P_{0}({\mathbf{r,r'}})\frac{\partial w_A({\mathbf{r}}^{\prime},t)}{\partial t}d{\mathbf{r}}^{\prime 3}= D\int_VP_{0}({\mathbf{r}},{\mathbf{r}}^{\prime})\nabla_{{\mathbf{r}}^{\prime}}^2\mu_A({\mathbf{r}}^{\prime})d{\mathbf{r}}^{\prime 3}
\end{equation}
leading to an equation of motion for the external fields:
\begin{equation}
\frac{\partial w_{A}(\mathbf{r})}{\partial t}=-D\nabla^2\mu_{A}(\mathbf{r})
\end{equation} 
Approximation (\ref{eq:approx_EPD}) is obviously exactly valid for a homogeneous mixture, because the pair--correlation 
function only depends on the distance $|{\mathbf{r}}-{\mathbf{r}}^{\prime}|$ between two points.  
In the inhomogeneous case, if ${\mathbf{r}}$ and ${\mathbf{r}}^{\prime}$ are in different phases and neither of
them in the interface, $\nabla_{{\mathbf{r}}}P_{0}({\mathbf{r,r'}})$ and 
$\nabla_{\mathbf{r'}}P_{0}(\mathbf{r,r'})$ are of different sign making this 
approximation justifiable even if the actual values differ.

Because of the incompressibility constraint $\mu_A$ and $\mu_B$ are not independent 
of each other. There is only one independent chemical potential $\mu=\mu_A-\mu_B$ so
the equation of motion for the external fields which is to be used in our case has
the form:
\begin{equation}
\frac{\partial w_{A}(\mathbf{r})}{\partial t}-\frac{\partial w_{B}(\mathbf{r})}{\partial t}=
-D\nabla^2\mu(\mathbf{r})
\label{eq:Maurits_EPD}
\end{equation}
If we compare this equation with equation (\ref{eq:diff_eq_EPD}) we see that  densities
evolving in time according to Rouse dynamics are well described through the EPD method if a 
local kinetic coefficient as given in Eq.(\ref{eq:Onsager_EPD}) is used.
\section{Random phase approximation for the fluctuations in EPD}
\label{app:RPA_EPD}
The single chain partition function $Q$ is defined through (compare with 
Eq.(\ref{eq:si_ch_part})):
\begin{equation}
Q=\int{\mathcal{D}}[{\mathbf{r}}_1]{\mathcal{P}}_1[{\mathbf{r}}]
\exp\left[-\frac{\rho}{N}\int_V d^3{\mathbf{r}}\;W({\mathbf{r}})
\hat{\phi}_1({\mathbf{r}})\right]
\label{eq:si_ch_part_2}
\end{equation}
where $\hat \phi_1({\bf r})=N/\rho \int {\rm d}s\; \delta({\bf r}-{\bf r}(s))$
denotes the single chain density. We now expect the system to be only weakly inhomogeneous,
i.e., meaning the density
and the external field only differ a little from the average value:
$\hat{\phi}_1({\mathbf{r}})=N/\rho V +\delta\phi_1({\mathbf{r}})$; 
$W({\mathbf{r}})=\bar{W}+\delta W({\mathbf{r}})$. 
The density and the external field are presented as a Fourier expansion:
\begin{equation}
\hat{\phi}_1({\mathbf{r}})=N/\rho V+\sum_{{\mathbf{q}}\neq 0}\hat\phi_{\mathbf{q}}
e^{i\mathbf{qr}}\qquad\qquad 
\hat\phi_{\mathbf{q}}=\frac{1}{V}\int_V d^3{\mathbf{r}}\;\hat{\phi}_1({\mathbf{r}})
e^{-i\mathbf{qr}}
\end{equation}
This Fourier expansion is now inserted in Eq.(\ref{eq:si_ch_part_2}):
\begin{equation}
\begin{split}
Q&=\int{\mathcal{D}}[{\mathbf{r}}_1]{\mathcal{P}}_1[{\mathbf{r}}]
\exp\left[-\bar{W}-\frac{\rho V}{N}\sum_{{\mathbf{q}}\neq 0}
W_{{\mathbf{q}}}\hat\phi_{-\mathbf{q}}\right]\\
&=\exp\left[-\bar{W}_A\right]\int{\mathcal{D}}[{\mathbf{r}}_1]
{\mathcal{P}}_1[{\mathbf{r}}_1]
\left\{1-\frac{\rho V}{N}\sum_{{\mathbf{q}}\neq 0}W_{{\mathbf{q}}}
\hat\phi_{-\mathbf{q}}+\frac{\rho^2V^2}{2N^2}\sum_{{\mathbf{q,q'}}\neq 0}
W_{{\mathbf{q}}}W_{{\mathbf{q'}}}\hat\phi_{-\mathbf{q}}
\hat\phi_{-\mathbf{q'}}+\cdots\right\}\\
&=\exp\left[-\bar{W}_A\right]Q_{0}\left\langle 1-
\frac{\rho V}{N}\sum_{{\mathbf{q}}\neq 0}W_{{\mathbf{q}}}
\hat\phi_{-\mathbf{q}}+\frac{\rho^2V^2}{2N^2}\sum_{{\mathbf{q,q'}}\neq 0}
W_{{\mathbf{q}}}W_{{\mathbf{q'}}}\hat\phi_{-\mathbf{q}}
\hat\phi_{-\mathbf{q'}}+\cdots\right\rangle_0
\end{split}
\end{equation}
$Q_{0}$ denotes the partition function of a single chain without an external 
field. The average in the last line is to be taken over all chain
configurations that are possible when there is no external field present. 
Because the average deviation of the density $\langle\delta\phi\rangle_0$ 
from the average value is zero if there is no external field, 
$\langle\phi_{\mathbf{q}}\rangle_0=0$ is valid. The average  
$\langle\phi_{-\mathbf{q}}\phi_{-\mathbf{q'}}\rangle_0$ is given through the single
chain structure factor $S_0({\mathbf{q}})$ of a Gaussian chain: $\langle\phi_{-\mathbf{q}}
\phi_{-\mathbf{q'}}\rangle_0=\frac{N}{\rho^2V^2}S_0({\mathbf{q}})\delta_{{\mathbf{-q,-q'}}}$. 
Neglecting higher terms, we obtain the RPA-result for the single chain partition function:
\begin{equation}
Q^{\text{RPA}}=\exp\left[-\bar{W}_A\right]Q_{0}
\exp\left[\frac{1}{2N}\sum_{{\mathbf{q}}\neq 0}S_0({\mathbf{q}})
|W_{{\mathbf{q}}}|^2\right]
\label{eq:q_rpa}
\end{equation}
Obviously a corresponding expression is valid for a B polymer. These RPA single
chain partition functions are plugged into Eq.(\ref{eq:free_en_UW}) 
leading to:
\begin{multline}
\frac{G\left[U,W\right]}{k_BT}= 
-\frac{\bar{\phi_A}\rho V}{N}\ln\frac{Q_{A,0}}{n_A}
-\frac{\bar{\phi_B}\rho V}{N}\ln\frac{Q_{B,0}}{n_B}
+\frac{\rho V \chi}{4}
+\frac{(\bar{\phi_A}-\bar\phi_B)\rho V \bar W}{2N}
+\frac{\rho V}{4N\chi N}\bar{W}^2+\\
+\frac{\rho V}{4N\chi N}\sum_{{\mathbf{q}}\neq 0}|W_{{\mathbf{q}}}|^2
-\frac{\bar{\phi_A}\rho V}{2N^2}
\sum_{{\mathbf{q}}\neq 0}S_0({\mathbf{q}})\left|\frac{(W_{{\mathbf{q}}}+U_{{\mathbf{q}}})}{2}\right|^2
-\frac{\bar{\phi_B}\rho V}{2N^2}
\sum_{{\mathbf{q}}\neq 0}S_0({\mathbf{q}})\left|\frac{(U_{{\mathbf{q}}}-W_{{\mathbf{q}}})}{2}\right|^2
\end{multline}
Regarding only the wave vector dependent parts of the free energy we find:
\begin{equation}
\begin{split}
\frac{G\left[U,W\right]}{k_BT}=&-\frac{\rho V}{2N^2}\sum_{{\mathbf{q}}\neq 0}
\left\{    \left(\frac{S_0({\mathbf{q}})}{4}-\frac{1}{2\chi}\right)|W_{{\mathbf{q}}}|^2
          +\left(\frac{S_0({\mathbf{q}})}{4}\right)|U_{{\mathbf{q}}}|^2\right\}+\\
&\qquad-\frac{\rho V}{2N^2}\sum_{{\mathbf{q}}\neq 0}\left\{
\frac{\bar{\phi}_A-\bar{\phi}_B}{2}S_0({\bf q})(U_{{\mathbf{q}}}W_{-{\mathbf{q}}})
\right\}+G_{\text{hom}}=\\
=&-\frac{\rho V}{2N^2}\sum_{{\mathbf{q}}\neq 0}
\left\{\left(\frac{S_0({\mathbf{q}})}{4}-\frac{1}{2\chi}-\frac{1}{4}(\bar{\phi}_A-\bar{\phi}_B)^2S_0({\mathbf{q}})
\right)|W_{\mathbf{q}}|^2\right\}+\\
&\qquad-\frac{\rho V}{2N^2}\sum_{{\mathbf{q}}\neq 0}
\left\{\frac{S_0({\bf q})}{4}\left|U_{{\mathbf{q}}}+(\bar{\phi}_A-\bar{\phi}_B)W_{{\mathbf{q}}}\right|^2\right\}+G_{\text{hom}}
\end{split}
\end{equation}
We use this free energy  to evaluate the partition function of
Eq.(\ref{eq:part_UW}). Following the procedure we used before, we employ a saddle
point approximation with respect to the field $U$:
\begin{equation}
\frac{\delta F[U,W]}{\delta U}|_{U^*}=0\qquad:\qquad U^*_{{\mathbf{q}}}= - (\bar{\phi}_A-\bar{\phi}_B)W_{{\mathbf{q}}}
\end{equation}
Since the free energy is quadratic in the RPA approximation the saddle point integration is equivalent to 
the functional integration over $U$. This again leads us to an expression for the free energy $G[W]$ only 
depending on the external field variable $W$:
\begin{equation}
\frac{G\left[W\right]}{k_BT}=\frac{\rho V}{2N^2}\sum_{{\mathbf{q}}\neq 0}
\left(\bar{\phi}_A\bar{\phi}_B S_0({\bf q}) -\frac{1}{2\chi} \right)|W_{{\mathbf{q}}}|^2+G_{\text{hom}}
\end{equation}
Now the average $\langle|W_{{\mathbf{q}}}|^2\rangle$ can 
be calculated:
\begin{equation}
\langle|W_{{\mathbf{q}}}|^2\rangle
=\frac{N^2}{\rho V}\frac{2\chi}{1-2\chi\bar{\phi}_A\bar{\phi}_BS_0({\mathbf{q}})}
\end{equation}
Using our result (\ref{eq:phi^2_av_Four}) 
\begin{equation}
\langle|\hat\phi_{{A\mathbf{q}}}-\hat\phi_{B{\bf q}}|^2\rangle=-\frac{4}{\rho V\chi}+
\frac{1}{(\chi N)^2}\langle|W_{{\mathbf{q}}}|^2\rangle
\end{equation}
we express the density fluctuations in terms of the field fluctuations
and recover the well--known RPA expression\cite{deGennes}:
\begin{equation}
\langle|\hat\phi_{{A\mathbf{q}}}-\hat\phi_{B{\bf q}}|^2\rangle=\frac{4}{\rho V}\left[
\frac{1}{\bar{\phi}_AS_0({\mathbf{q}})}+\frac{1}{\bar{\phi}_BS_0({\mathbf{q}})}
-2\chi\right]^{-1}  \equiv 4 S_{\rm RPA}({\bf q})
\end{equation}
Using the RPA single chain partition function (\ref{eq:q_rpa}), we calculate
\begin{eqnarray}
\phi^*_A({\bf r}) &=& - \frac{\bar \phi_A V}{Q_A} \frac{\delta Q_A}{\delta W_A({\bf r})}
    = \bar \phi_A - \frac{\bar \phi_A}{N} \sum_{{\bf q}\neq 0} S_0({\bf q}) W_{A{\bf q}}e^{i{\bf qr}} \\
\frac{\delta\phi^*_A({\bf r})}{\delta W_A({\bf r}')} &=& - \frac{\bar \phi_A}{NV} \sum_{{\bf q}\neq 0} S_0({\bf q}) e^{i{\bf q}({\bf r}-{\bf r}')}\label{eq:pair-corr}
\end{eqnarray}
The last equation is equivalent to Eq.\ref{eq:pair-corr_EPD} but due to the 
particle conservation the ${\bf q}\!=\!0$ contribution has to be taken
out of the sum. $\phi_A({\bf q}=0)$ is just the average overall density
${\bar{\phi}}_A$ that cannot change if the external field is altered.
With these expressions we obtain for the ``literal'' fluctuations in the EPD method according to Eq.(\ref{eq:phi^2_av_ok}):
\begin{eqnarray}
\langle|\hat \phi_{{A\mathbf{q}}}-\hat \phi_{B{\bf q}}|^2\rangle_{\rm EPD} &=&
     \langle|\phi^*_{{A\mathbf{q}}}-\phi^*_{B{\bf q}}|^2\rangle - \frac{N}{\rho V^2} \int {\rm d}^3{\bf r}\;{\rm d}^3{\bf r}'\; e^{i {\bf q}({\bf r}-{\bf r}')}
     \left\langle \frac{\delta \phi^*_A({\bf r})}{\delta W_A({\bf r}')} + \frac{\delta \phi^*_B({\bf r})}{\delta W_B({\bf r}')}\right\rangle  \nonumber \\
     &=& \frac{8\chi S_0^2({\bf q})\bar\phi_A^2\bar\phi_B^2}{\rho V(1-2\chi\bar \phi_A \bar \phi_B S_0({\bf q}))} + \frac{S_0({\bf q})}{\rho V} \nonumber \\
     &=&  4 S_{\rm RPA}({\bf q}) + \frac{(1-4\bar \phi_A \bar \phi_B)S_0({\bf q})}{\rho V}
\end{eqnarray}
Generally, the deviation for the RPA result is of similar magnitude as the RPA structure factor itself. For a symmetric quench $\bar \phi_A = \frac{1}{2}$, however, 
we accidentally recover the RPA result.  This example also illustrates that one can obtain the average of the composition by sampling the average of 
$\phi_A^*-\phi_B^*$, i.e., the densities of single chains in the field configuration $W$, but one should not use this to calculate fluctuations.



\begin{thebibliography}{10}
\bibitem{COMMENT}
This is common practice in self consistent field calculations. It is understood
  that the incompressibility constraint is enforced only after coarse--graining
  the microscopic densities over a small but finite volume. On smaller length
  scales the specific spatial dependence of the repulsive interactions matters
  and leads, e.g., to packing effects. This is beyond our treatment.
\end{thebibliography}
\end{document}